\documentclass[12pt,draftcls,onecolumn]{IEEEtran}
\usepackage{amsmath,amssymb,graphicx,epstopdf,cite,subfigure,multirow,array}
\usepackage{psfrag}
\usepackage{amssymb}
\usepackage{amsmath}
\usepackage{bbm}
\usepackage{dsfont}
\usepackage{pifont}
\usepackage{cite}
\usepackage{graphics}
\usepackage{graphicx}
\usepackage{epsfig}
\usepackage{subfigure}
\usepackage{amscd}
\usepackage{accents}

\newtheorem{theorem}{Theorem}

\newtheorem{definition}[theorem]{Definition}

\newlength{\dhatheight}
\newcommand{\doublehat}[1]{%
    \settoheight{\dhatheight}{\ensuremath{\hat{#1}}}%
    \addtolength{\dhatheight}{-0.2ex}%
    \hat{\vphantom{\rule{1pt}{\dhatheight}}%
    \smash{\hat{#1}}}}

\title{Sparse Detection of Non-Sparse Signals for Large-Scale Wireless Systems}

\author{
Jun Won Choi,~\IEEEmembership{Member,~IEEE,}
        and Byonghyo~Shim,~\IEEEmembership{Senior Member,~IEEE}
        \thanks{J. Choi is with Dept. of Electrical Engineering, Hanyang University, Korea and B. Shim is with Dept. of Electrical and Computer Engineering, Seoul National University, Korea.}

\thanks{This research was funded by the research grant from Qualcomm Incorporated and National Research Foundation of Korea (NRF) funded by the Korea government (MEST) (No. 2012R1A2A2A01047510). }
\thanks{This paper was presented in part at the Globecom conference, Austin, Dec. 2014.}
}

\begin{document}

\maketitle

\begin{abstract}
In this paper, we introduce a new detection algorithm for large-scale wireless systems, referred to as post sparse error detection (PSED) algorithm, that employs a sparse error recovery algorithm to refine the estimate of a symbol vector obtained by the conventional linear detector. The PSED algorithm operates in two steps: 1) sparse transformation converting the original non-sparse system into the sparse system whose input is an error vector caused by the symbol slicing and 2) estimation of the error vector using the sparse recovery algorithm. From the asymptotic mean square error (MSE) analysis and empirical simulations performed on large-scale systems, we show that the PSED algorithm brings significant performance gain over classical linear detectors while imposing relatively small computational overhead.
\end{abstract}

\begin{IEEEkeywords}
Sparse signal recovery, compressive sensing, large scale systems, orthogonal matching pursuit, sparse transformation, linear minimum mean square error, error correction.
\end{IEEEkeywords}

\IEEEpeerreviewmaketitle

\section{Introduction}
As a paradigm guaranteeing the perfect reconstruction of a sparse signal from a small set of linear measurements, compressive sensing (CS) has generated a great deal of interest in recent years.
Basic premise of the CS is that the sparse signals $\mathbf{x} \in \mathbb{R}^{n_r}$ can be reconstructed from the compressed measurements $\mathbf{y} = \mathbf{H} \mathbf{x} \in \mathbb{R}^{n_t}$ ($n_t < n_r$) as long as the signal to be recovered is sparse (i.e., number of nonzero elements in the vector is small) and the measurement process approximately preserves the energy of the original sparse vector \cite{cs_magazine,donoho}.
The CS paradigm works well in many signal processing applications where the signal vector to be reconstructed is sparse in itself or sparse in a transformed domain.
In particular, CS technique has been applied to wireless communication applications in the context of sparse channel estimation \cite{preisig,berger} and wireless multiuser detection \cite{zhu,bshim_mud,dekorsy}, where the multi-dimensional quantities being estimated exhibit sparsity structure. However, not much work is available for the information vector detection (possible exception can be \cite{bshim_sparse}) mainly because the information vectors being transmitted in a typical communication system are by no means sparse so that the sparse recovery algorithm would not outperform the conventional receiver algorithm, not to mention having no performance guarantee.

It is to these types of wireless detection problem that this paper is addressed. This problem,
which is seemingly unconnected to the CS principle, is prevalent and embraces many of current
and future detection problems in wireless communication scenarios including massive multiple-input-multiple-output (MIMO), internet of things (IoT), multiuser detection, interference cancellation, source localization, to name just a few.
Traditional way of detecting the input signals is classified into two categories: linear detection and nonlinear detection techniques. Linear detection techniques, such as zero forcing (ZF) or linear minimum mean square error (LMMSE) estimation, are simple to implement and easy to use but the performance is not appealing when compared to the nonlinear detectors [9]. Nonlinear detection schemes usually perform better than the linear detection but it requires significant computational overhead. Recently, several suboptimal detection algorithms such as the K-best algorithm \cite{kbest} and the fixed-complexity sphere decoder \cite{fsd} have been proposed, but still these approaches are computationally challenging in the detection of large dimensional systems.

%
%

Our approach provides a new solution to the large-scale detection problems by deliberately combining the linear detection and a (nonlinear) sparse signal recovery algorithm with the aim of improving the receiver performance while keeping the computational complexity low. Our proposed algorithm, henceforth dubbed as post sparse error detection (PSED), is based on the simple observation that the conventional linear detection algorithm performs reasonably well and thus the error vector after the slicing (symbol quantization) of the detector output is well modeled as a sparse vector.
In a nutshell, the PSED algorithm operates in two steps.
In the first step, we perform the conventional linear detection to generate a {\it rough} estimate of the transmit symbol vector. Since the performance of conventional detector is acceptable in the operating regime, the error vector obtained by the slicing of detected symbol vector is readily modeled as a sparse signal.
Now, by a simple linear transform of this error vector, we can obtain the new measurement vector whose input is the {\it sparse error vector}. In the second step, we use the sparse recovery algorithm to estimate the sparse error vector and then cancel it from the sliced symbol vector, the sum of the original symbol vector and the error vector. As a result of this error cancellation, we obtain more reliable estimate of the original symbol vector.

In our random matrix based analysis, we show that the asymptotic performance of the proposed PSED algorithm, measured in terms of mean square error (MSE), decays exponentially with signal-to-noise ratio (SNR), which is in contrast to the linear or sublinear decaying behavior of the conventional linear detectors.
In fact, we show from the empirical simulations that the PSED scheme outperforms the conventional linear detectors by a large margin and thus performs close to maximum likelihood detection algorithm.

The rest of this paper is organized as follows. In Section~\ref{sec:algorithm}, we describe the system model and the proposed PSED algorithm. In Section~\ref{sec:recovery}, we introduce the CS recovery algorithm for the PSED technique. In  Section~\ref{sec:analysis}, the asymptotic performance analysis of the PSED scheme is provided using random matrix theory. In Section~\ref{sec:simulation}, we present the simulation results and conclude the paper in Section~\ref{sec:conclusion}.

\section{Post Sparse Error Detection}
\label{sec:algorithm}

\subsection{System Model and Conventional Detectors}
The relationship between the transmit symbol and the received
signal vector in many wireless systems can be
expressed as
\begin{eqnarray}
\label{eq:model1}
\mathbf{y} = \sqrt{P} \mathbf{H} \mathbf{s} + \mathbf{v}
\end{eqnarray}
where  $\mathbf{y} \in \mathbb{C}^{n_r}$ is the received signal vector, $\mathbf{s} \in \mathbb{C}^{n_t}$ is the transmit symbol vector whose entries are chosen from a set $\Omega$ of finite symbol alphabet, $\mathbf{H} \in \mathbb{C}^{n_r \times n_t}$ is the channel matrix,  $\mathbf{v}\sim \mathcal{CN}(0,\sigma_{v}^{2}\mathbf{I}_{n_r})$ is the noise vector, and $P$ is the transmitted power.
%
%
In detecting the transmit information from the received signals, we have two options: linear detection and nonlinear detection schemes.
In the linear detection scheme, an estimate $\tilde{\mathbf{s}}$ of the transmit symbol vector is obtained by applying the weight matrix $\mathbf{W} \in \mathbb{C}^{n_r \times n_t}$ to the received vector $\mathbf{y}$
\begin{align} \label{eq:lin}
\tilde{\mathbf{s}} = \mathbf{W}^{H} \mathbf{y}.
\end{align}
The well-known weight matrices for the linear detector $\mathbf{W}$ include \cite{verdu}
\begin{itemize}
\item Matched filter (MF): $\mathbf{W} = \frac{1}{\sqrt{P}}\mathbf{H}$
\item ZF receiver: $\mathbf{W} = \frac{1}{\sqrt{P}} \mathbf{H} \left(\mathbf{H}^{H}\mathbf{H}\right)^{-1}$
\item LMMSE receiver: $\mathbf{W} = \mathbf{H} \left(\mathbf{H}^{H}\mathbf{H}+\frac{\sigma_{v}^{2}}{P}\mathbf{I}\right)^{-1}$.
\end{itemize}
The linear detection is simple to implement and computationally efficient, but the performance is typically not better than the nonlinear detection scheme.
 The nonlinear detectors, such as ML and maximum a posteriori (MAP) detectors, exploit the additional side information that an element of the transmit vector is chosen from the set of finite alphabets.
For the lattice that the symbol vector spans, a search is performed to find out a solution minimizing the cost function. In the ML detector, for example, a symbol vector ${\mathbf{s}}$ minimizing the ML cost function $J(\mathbf{s}) = \min_{\mathbf{s} \in \Omega^{n_t}} \| \mathbf{y}-\sqrt{P}\mathbf{H}\mathbf{s}\|^{2}$ is chosen among all possible candidates.
When compared to the linear detection schemes, the nonlinear detector offers better performance but it requires higher computational cost.
Sphere decoding (SD) algorithm, for example, performs an efficient ML detection using the closest lattice point search (CLPS) in a hypersphere with a small radius \cite{fincke}.
In spite of the substantial reduction in complexity over the brute force enumeration scheme, computational burden of the SD algorithm is still a major problem, since the expected complexity is exponential with the problem size \cite{Jalden}.
%
%
%
Due to these reasons, in many future wireless scenarios where the dimension of the system matrix is much larger than that of today's systems, both linear and nonlinear principles have their own drawback and may not offer an elegant tradeoff between performance and complexity.

\subsection{Sparse Transform via Conventional Detection}
\label{sec:sparse_transform}
%
When we apply conventional detectors to the system of \eqref{eq:model1}, it is clear that the detector output is similar but not always identical to the original information vector $\mathbf{s}$.
For a practical SNR regime, therefore, the detector output might contain an error, causing mismatches for a few entries of $\mathbf{s}$.\footnote{In the operating regime of communication systems, symbol error rate (SER) is typically less than $10\%$.}
Our approach is to exploit such sparse nature of the detection errors.
To leverage the sparsity of the detection errors, we need to somehow convert the non-sparse system into the sparse one. Conventional detection together with the symbol slicing serves our purpose since the estimated symbol vector is {\it roughly} accurate, and hence, the resulting error vector (defined as the difference between the original symbol vector and the sliced estimate) is well modeled as a sparse signal.
Denoting the estimate of a symbol vector as $\tilde{\mathbf{s}}$ and its sliced version as $\hat{\mathbf{s}}$, one can express $\hat{\mathbf{s}}$ as
\begin{eqnarray}
    \hat{\mathbf{s}} =  Q(\tilde{\mathbf{s}}) = \mathbf{s} -  \mathbf{e} \label{eq:err}
\end{eqnarray}
where $Q(\cdot)$ is the slicing function and $\mathbf{e}$ is the error vector.
%
%
%
As mentioned, in an operational regime of communication systems, the number of nonzero entries (i.e., {\it real errors}) in $\mathbf{e}$ would be small so that the error vector is well modeled as a sparse vector.
Suppose the dimension of the symbol vector $\mathbf{s}$ is $16$ and the symbol error rate is $10\%$, then the probability that more than $5$ elements are in error is $0.3\%$ while that of $5$ or less elements being in error are $99.7\%$.
As long as the error vector is sparse, by transmitting this error vector, one can construct a sparse system expressed in terms of the error vector $\mathbf{e}$. This task, henceforth referred to as the {\it sparse transform}, is realized by the re-generation of the received signal from the detected symbol $\hat{\mathbf{s}}$ followed by the subtraction as
\begin{equation} \label{eq:stransform}
     \mathbf{y}' = \mathbf{y} - \sqrt{P} \mathbf{H}  \hat{\mathbf{s}},
\end{equation}
where $\mathbf{y}'$ is the newly obtained received vector (see Fig. \ref{fig:proposed}). Then, from (\ref{eq:model1}), (\ref{eq:err}) and (\ref{eq:stransform}), the new measurement vector $\mathbf{y}'$ is written by
\begin{align}
\mathbf{y}'  &=  \sqrt{P} \mathbf{H} (\mathbf{s} - \hat{\mathbf{s}}) + \mathbf{v} \nonumber \\
&= \sqrt{P} \mathbf{H} \mathbf{e} + \mathbf{v}. \label{eq:stransform2}
\end{align}

Interestingly, by adding trivial operations (matrix-vector multiplication and subtraction), we can convert the original non-sparse system into the sparse system whose input is an error vector associated with the conventional detector.
Note that the symbol slicing is essential in sparsifying the error vector and we have two options:
\begin{itemize}
\item Hard-slicing: hard-slicing literally performs the hard decision of symbol estimate. The slicer function maps the input to the closest value in the symbol set $\Gamma$ (i.e., $Q(z) = \arg \min_{\gamma \in \Gamma} \| z - \gamma \|_2$). By exploiting the discrete property of the transmit vectors, we can enforce the sparsity of the input (error vector) for the modified system.
    Main benefit of the hard slicing is that it entirely removes the residual interference and noise when the estimate lies in the decision region of the original symbol.
\item Soft-slicing: when the {\it a prior} information on the source exists, soft slicing might be a useful option.
      One possible way is to use an MMSE-based soft slicing ($\hat{s}_i = E [s_i | \tilde{s}_i]$), where $s_i$ denotes the $i$-th element of $\mathbf{s}$ \cite{shim_choi_kang}.
     For example, when the decoder feeds back the log-likelihood ratio (LLR) information $L_b$ on the information bit, one can transform this into the symbol prior information $P_r (s_i)$.
    %
Using the nonuniform symbol probability $P(s_i)$, better estimate $\hat{s}_i$ is obtained as
        \begin{align}
        \label{eq:slice7}
        \hat{s}_i = E [s_i | \tilde{s}_i] = \sum_{s_i \in \Omega} s_i  P_r (s_i | \tilde{s}_i )
                  =  \frac {  \sum_{s_i \in \Omega} s_i  P_r ( \tilde{s}_i | s_i ) P_r ( s_i ) }{\sum_{s_i \in \Omega}  P_r ( \tilde{s}_i | s_i ) P_r ( s_i ) }  .
        \end{align}
When the linear detectors in (\ref{eq:lin}) are used to obtain $\tilde{\mathbf{s}}$, $P_r ( \tilde{s}_i | s_i )$ is given by
        \begin{align}
        P_r ( \tilde{s}_i | s_i ) &= \frac{1}{2\pi \sigma_s^2} \exp\left(-\frac{1}{\sigma_s^2} \left( \hat{s}_{i} - \sqrt{P} \mathbf{w}_{i}^{H}\mathbf{h}_{i}s_i  \right)^{2} \right),
              \end{align}
        where
          $\sigma_{s}^{2} = \mathbf{w}_{i}^{H}\left(P \mathbf{H}\mathbf{H}^{H}+\sigma_v^2 \mathbf{I} \right)\mathbf{w}_i - P(\mathbf{w}_{i}^{H}\mathbf{h}_{i})^{2}$,
        and $\mathbf{w}_{i}$ and $\mathbf{h}_{i}$ are the $i$-th column of $\mathbf{W}$ and $\mathbf{H}$, respectively.
      Although the soft slicing does not strictly enforce the sparsity of the resulting system, it provides better shaping of the symbol so that the number of {\it nontrivial} nonzero elements (errors with large magnitude) in the error vector $\mathbf{e}$ can be reduced.
\end{itemize}
In case of hard slicing, all entries of $\mathbf{e}$ are zero except for those associated with the detection errors. On the other hand, when the soft-slicing is employed, the entries unassociated with the detection errors might have small nonzero magnitude, yielding so called approximately sparse vector $\mathbf{e}$\footnote{Approximately sparse signal is referred to as the one that contains most of energy in only a few coefficients.}.
%
In our analysis and derivation that follow, we will mainly focus on the hard-slicing based PSED technique for analytical simplicity.

\subsection{Recovery of Sparse Error Vector}
\label{sec:error_recovery}
\begin{figure}[t]
\begin{center}
\ifCLASSOPTIONonecolumn
	\includegraphics[width=140mm]{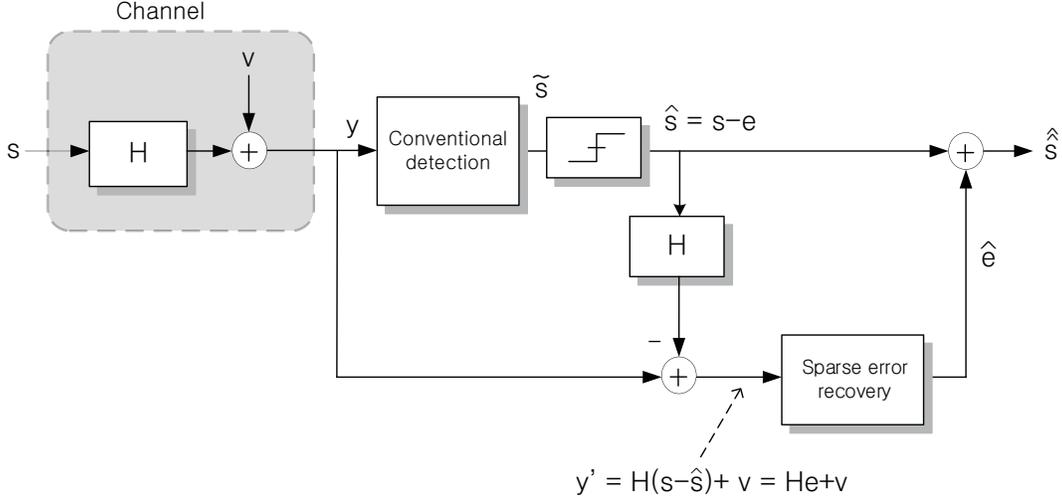}
\else
	\includegraphics[width=85mm]{sparse_det_model6}
\fi
\caption{Overall structure of the proposed PSED detection algorithm.}
\label{fig:proposed}
\end{center}
\end{figure}

%
Once the non-sparse system is converted into the sparse one, we can use the sparse recovery algorithm to estimate the error vector. To be specific, using newly obtained measurement vector $\mathbf{y}'$ and the channel matrix $\mathbf{H}$, sparse recovery algorithm estimates the error vector $\mathbf{e}$ (see Fig. \ref{fig:proposed}). There are many algorithms designed to recover the sparse vector in the presence of noise. Well-known examples include basis pursuit de-noising (BPDN) \cite{bpdn} and orthogonal matching pursuit (OMP) \cite{omp}. We will say more about this in Section \ref{sec:recovery}.
Once the output $\hat{\mathbf{e}}$ of the sparse recovery algorithm is obtained, we add it to the sliced detector output $\hat{\mathbf{s}}$, generating the refined symbol vector $\doublehat{\mathbf{s}}$
%
\begin{eqnarray}
\doublehat{\mathbf{s}} = \hat{\mathbf{s}} + \hat{\mathbf{e}} = (\mathbf{s} - \mathbf{e}) + \hat{\mathbf{e}} = \mathbf{s} + (\hat{\mathbf{e}} - \mathbf{e}).  \label{eq:err_correct}
\end{eqnarray}
%
If the error estimate is accurate, i.e., $\hat{\mathbf{e}} \approx \mathbf{e}$, then the magnitude of the error difference $\epsilon = \hat{\mathbf{e}} - \mathbf{e}$ would be small so that the re-estimated symbols $\doublehat{\mathbf{s}}$ becomes more accurate than the initial estimate $\hat{\mathbf{s}}$.
As long as the sparsity of the error vector is ensured (i.e., number of nonzero elements in error vector $\mathbf{e}$ is small), an output of the sparse recovery algorithm $\hat{\mathbf{e}}$ would be faithful and hence the refined symbol vector will be more reliable than the original estimate (i.e., $E\|\hat{\mathbf{e}} -  \mathbf{e} \|_2^2 < E\|  \mathbf{e} \|_2^2$).
This can be easily explained for noiseless scenario; if we can identify the support (index set of nonzero entries) of the error vector via the sparse recovery algorithm such as OMP, we can convert the underdetermined system into the overdetermined system by removing columns associated with the zero element in $\mathbf{e}$. Since the LS estimate reconstructs the original symbol vector accurately ($\hat{\mathbf{e}} = \mathbf{e}$), we have $E\|\hat{\mathbf{e}} - \mathbf{e} \|_2^2 = 0 < E\|  \mathbf{e} \|_2^2$. This argument, however, does not hold true for noisy scenario and we need more deliberate analysis (see Section \ref{sec:analysis}).

\begin{table}[t]
\begin{center}
\caption{Operations of the PSED detector}
\begin{tabular}{ll}
\hline
Input: & $\mathbf{y},\mathbf{H}$ $\;\;\;\;\;$ Output: $\hat{\mathbf{s}}_{\rm final}$\\
\hline\hline
Step 1: & Perform conventional detection to obtain $\tilde{ \mathbf{s} }$. \\
Step 2: & Perform sparse transform, i.e.,  $\mathbf{y}' = \mathbf{y} - \sqrt{P} \mathbf{H} \hat{\mathbf{s}}$ where $\hat{\mathbf{s}} = Q( \tilde{ \mathbf{s} } )$. \\
Step 3: & Apply the sparse recovery algorithm to $\mathbf{y}'$ to estimate $\hat{\mathbf{e}}$.  \\
Step 4: & Correct the detection errors in $\hat{\mathbf{s}}$, i.e.,  $\doublehat{\mathbf{s}} = \hat{\mathbf{s}} + \hat{\mathbf{e}}$. \\
Step 5: & Generate the final symbol estimate, i.e.,   $\hat{\mathbf{s}}_{\rm final} = Q(\doublehat{\mathbf{s}})$.  \\
\hline
\label{tb:proc_pdsr}
\end{tabular}
\end{center}
\end{table}
%
%

\begin{table}[t]
\begin{center}
\caption{Similarities between the decoding process of the linear block code and the PSED algorithm. In the decoding process, $\mathbf{r}$, $\mathbf{c}$, and $\mathbf{s}$ denote received, codeword, and syndrome vectors, respectively, and $\mathbf{H}$ is the parity check matrix.}
\label{similar}
\begin{tabular}{l|l}
\hline
 Decoding process of the linear block code & Proposed PSED algorithm   \\
\hline \hline
      Received vector: $\mathbf{r} = \mathbf{c} + \mathbf{e}$  & Detected symbol vector: $\hat{\mathbf{s}} = \mathbf{s} - \mathbf{e}$ \\ \cline{1-2}
      Syndrome vector: $\mathbf{s} = \mathbf{e}\mathbf{H}^T$          & New observation vector: $\mathbf{y}' = \sqrt{P}\mathbf{H e} + \mathbf{v}$ \\ \cline{1-2}
      Recovered codeword: $\hat{\mathbf{c}} = \mathbf{r} + \hat{\mathbf{e}}$  & Re-detected symbol vector: $\doublehat{\mathbf{s}} = \hat{\mathbf{s}} + \hat{\mathbf{e}}$ \\ \cline{1-2}
\end{tabular}
\end{center}
\end{table}

%
%
%
%
It is worth mentioning that the sparse error recovery process is analogous to the decoding process of the linear block code \cite{wicker}.
First, the sliced symbol vector $\hat{\mathbf{s}} = \mathbf{s} - \mathbf{e}$ can be viewed as a received vector (often denoted by $\mathbf{r}$ in the coding theory) which is expressed as the sum of the transmit codeword and the error vector.
Also, the new observation vector $\mathbf{y}' =\sqrt{P}\mathbf{H} \mathbf{e} + \mathbf{v}$ is similar in spirit to the syndrome, the product of the error vector and the transpose of parity check matrix.
Note that the syndrome is a sole function of the error vector and does not depend on the transmit codeword.
Similarly, the new observation vector $\mathbf{y}'$ is a function of $\mathbf{e}$ and independent of the transmit vector $\mathbf{s}$.
Furthermore, in the linear block code, the decoded error pattern is correct only when the cardinality of the error vector is within the error correction capability $t$ (i.e., $\| \mathbf{e} \|_0 < t$) and similar behavior occurs to the problem at hand since an output of the sparse recovery algorithm will be reliable only when the error vector $\mathbf{e}$ is sparse (i.e., $\| \mathbf{e} \|_0 \ll n_t$).
Finally, the error correction is performed in the decoding process by adding the reconstructed error vector $\hat{\mathbf{e}}$ and the received vector $\mathbf{r}$ and the same is true for the proposed algorithm (see \eqref{eq:err_correct}). In Table I and II, we summarize the proposed PSED algorithm and also compare the PSED algorithm and the decoding process of the linear block code.

\vspace{0.5cm}
\section{Sparse Error Vector Recovery} \label{sec:recovery}

\vspace{0.5cm}
%
%
In this section, we briefly describe the sparse recovery algorithm used for the proposed PSED scheme.
Note that an approach often called Lasso or Basis pursuit de-noising (BPDN) formulates the problem to recover the sparse signal in the noisy scenario as \cite{lasso}
\begin{gather}
\min_{\mathbf{e}} ~ \frac{1}{2} \| \mathbf{y}' - \sqrt{P}\mathbf{H} \mathbf{e} \|_2^2 + \lambda \| \mathbf{e} \|_1  \nonumber
\end{gather}
where $\lambda$ is the penalty term to control the amount of weight given to the sparsity of the desired signal $\mathbf{e}$.
 This problem is in essence a convex optimization problem and there are many algorithms to solve this type of problem \cite{boyd}. 
Recently, greedy algorithms have received much attention as an alternative for the convex optimization problem. In a nutshell, greedy algorithms attempt to find the support of $\mathbf{e}$ (i.e., the set of columns in $\mathbf{H}$ constructing $\mathbf{y}'$) in an iterative fashion, generating a sequence of the estimate for $\mathbf{e}$.
In the OMP algorithm, for example, a column of $\mathbf{H}$ maximally correlated with the modified measurements (residual $\mathbf{r}$) is chosen as an element of the support set $\hat{\mathcal{E}}$ \cite{omp}.
Then the estimate of the desired signal $\hat{ \mathbf{e} }_{\hat{\mathcal{E}}}$ is constructed by projecting $\mathbf{y}'$ onto the subspace spanned by the columns supported by $\hat{\mathcal{E}}$. That is,
\begin{eqnarray}
\hat{ \mathbf{e} }_{\hat{\mathcal{E}}}
&=& \label{eq:bff1} \frac{1}{\sqrt{P}} {{\left( \mathbf{H }_{  \hat{\mathcal{E}}   }^H{{\mathbf{H }}_{ \hat{\mathcal{E}}   }}
\right)}^{-1}}\mathbf{H }_{   \hat{\mathcal{E}}   }^H \mathbf{y}'
\end{eqnarray}
%
%
Finally, we update the residual $\mathbf{r}$ so that it contains measurement excluding those included by the estimated support set (${\mathbf{r}} = \mathbf{y}' - \sqrt{P} \mathbf{H }_{  \hat{\mathcal{E}} } \hat{ \mathbf{e} }_{\hat{\mathcal{E}}}$).
If the OMP algorithm identifies the support $\mathcal{E}$ accurately ($\hat{\mathcal{E}} = \mathcal{E}$), then one can remove all non-support elements in $\mathbf{e}$ and corresponding columns in $\mathbf{H}$ so that one can obtain the overdetermined system model
\begin{align}
\mathbf{y}' = \sqrt{P} \mathbf{H}_\mathcal{E} \mathbf{e}_\mathcal{E} + \mathbf{v}.
\label{eq:overdet}
\end{align}
%
%
and the final estimate is equivalent to the best estimate often called the Oracle LS estimator,\footnote{In case the {\it a prior} information on signal and noise is available, one can alternatively use the linear minimum mean square error (LMMSE) estimate. For example, if $\sigma_e^2 = E[ | \mathbf{e}_\mathcal{E} |^2]$ and $\sigma_v^2 = E[ | \mathbf{v} |^2]$, we have $\hat{\mathbf{e}}_\mathcal{E} = \frac{1}{\sqrt{P}} ( \mathbf{H}_\mathcal{E}^H \mathbf{H}_\mathcal{E} + \frac{ \sigma_{v}^2 }{P \sigma_{e}^2 } \mathbf{I})^{-1} \mathbf{H}_{\mathcal{E}}^{H} \mathbf{y}'$
}
\begin{equation}
\hat{\mathbf{e}}_\mathcal{E} =\frac{1}{\sqrt{P}}{\left(\mathbf{H}_{\mathcal{E}}^{H}\mathbf{H}_{\mathcal{E}}\right)^{-1}\mathbf{H}_{\mathcal{E}}^{H}}\mathbf{y}'.
\nonumber \label{eq:final_output}
\end{equation}
%
%
%
While the OMP algorithm is simple to implement and also computationally efficient, due to the selection of the single candidate in each iteration, the performance depends heavily on the selection of index.
In fact, the output of OMP would be wrong if a single incorrect index (index not contained in the support) is chosen in the middle of the search. In order to alleviate this drawback, various approaches investigating {\it multiple indices} have been suggested. Recent developments of this approach include the regularized OMP \cite{romp}, compressive sampling matching pursuit (CoSaMP) \cite{cosamp}, subspace pursuit (SP) \cite{Dai}, and generalized OMP \cite{gomp}.

\begin{figure*}[t]
	\begin{center}
		\includegraphics[height=60mm,width=140mm]{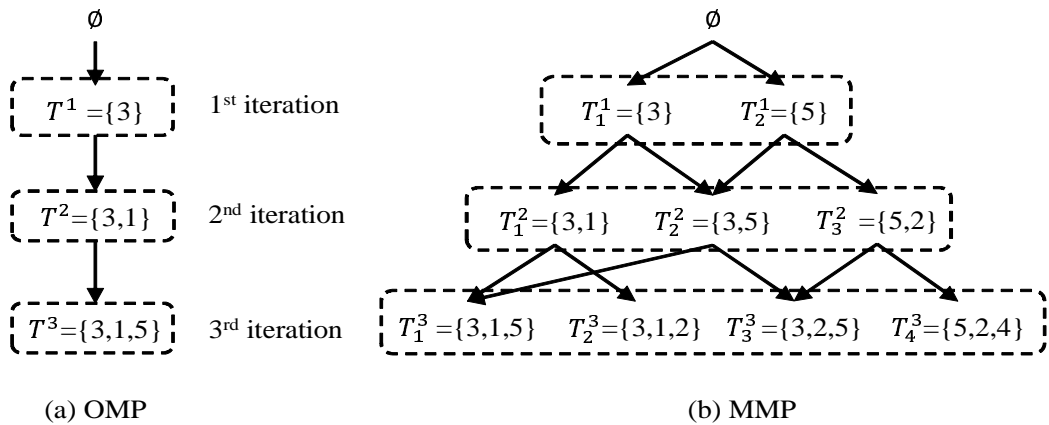}
		\caption{Comparison between the OMP and the MMP algorithm ($L=2$ and $K=3$). While OMP maintains a single candidate $T^k$ in each iteration, MMP investigates multiple promising candidates $T^k_j$ ($1 \leq j \leq L$) (the subscript $j$ counts the candidate in the $i$-th iteration).}
		\label{fig:concept_tree}
	\end{center}
\end{figure*}

\begin{table*}[!t]
\centering
\caption{The MMP algorithm}
\label{tab:mmp_alg}
\begin{tabular}{l r}
\hline
\hline
\multicolumn{2}{l}{\textbf{Input:} measurement $\mathbf{y}$, sensing matrix $\mathbf{\Phi}$, sparsity $K$, number of path $L$ }\\
\multicolumn{2}{l}{\textbf{Output:} estimated signal $\hat{\mathbf{e}}$	}\\
\multicolumn{2}{l}{\textbf{Initialization:} $k:=0$ (iteration index), $\mathbf{r}^0 := \mathbf{y}$ (initial residual), $S^0 := \left \{ \emptyset \right \}$		 }\\
\hline
\multicolumn{2}{l}{{\bf while} {$k < K$} {\bf do}										}\\
\multicolumn{2}{l}{	\hspace{4mm}$k :=k+1$, $u :=0$, $S^{k} :=\emptyset$				 }\\
\multicolumn{2}{l}{	\hspace{4mm}{\bf for} {$ i = 1$ {\bf to} $| S^{k-1} | $} {\bf do}			 }\\
					\hspace{8mm}	$ \tilde{\pi} := \arg \mathop {\max} \limits_{ \left | \pi \right | = L} { \| ( \mathbf{H}^{H} \mathbf{r}^{k-1}_i )_{\pi} \|_2^2 } $ & ({\it choose $L$ best indices})	\\
\multicolumn{2}{l}{	\hspace{8mm}	{\bf for} {$ j=1 $ {\bf to} $ L $} {\bf do}													 }\\
					\hspace{12mm}		$s_{tmp} := s^{k-1}_i \cup \left \{ \tilde{\pi}_j \right \}$	& ({\it construct a temporary path})\\
					\hspace{12mm}		{\bf if} {$s_{tmp} \not \in S^{k}$}	{\bf then} & ({\it check if the path already exists})		 \\
					\hspace{16mm}			$u :=u+1$							& ({\it candidate index update})		 \\
                    \hspace{16mm}			$s^k_u :=s_{tmp}$		& ({\it path update})
\\
					\hspace{16mm}			$S^{k} := S^{k} \cup \{ s^{k}_u \}$	& ({\it update the set of path})\\
					\hspace{16mm}			$\hat{\mathbf{e}}^k_u := \mathbf{H}^{\dagger}_{s^{k}_u} \mathbf{y}$	& ({\it perform estimation})	 \\
					\hspace{16mm}			$\mathbf{r}^k_u := \mathbf{y}-\mathbf{H}_{s^{k}_u} \hat{\mathbf{e}}^k_u$	 & ({\it residual update})\\
\multicolumn{2}{l}{	\hspace{12mm}		{\bf end if}																	 }\\
\multicolumn{2}{l}{	\hspace{8mm}	{\bf end for}																		 }\\
\multicolumn{2}{l}{	\hspace{4mm}{\bf end for}																			 }\\
\multicolumn{2}{l}{{\bf end while}																							 }\\
					{$ u^* := \arg \mathop {\min}_{u} { \| \mathbf{r}_{u}^K \|_2^2 } $}		& ({\it find index of the best candidate})		\\
\multicolumn{2}{l}{{$ s^* := s^K_{u^*}$}					 }\\
\multicolumn{2}{l}{{\bf return} $\hat{\mathbf{e}} = \mathbf{H}^{\dagger}_{s^*} \mathbf{y}$															 }\\
\hline
\end{tabular}
\end{table*}

In this work, we employ the multipath matching pursuit (MMP) algorithm, recently proposed greedy tree search algorithm, in recovering the sparse error vector \cite{mmp}.
While aforementioned greedy recovery algorithms identify the elements of support sequentially and choose a single support estimate $\hat{\mathcal{E}}$, MMP performs the parallel search to find the multiple promising supports (we henceforth refer to it as {\it support candidates}) and then chooses the best candidate minimizing the residual power in the last minute.
As shown in Fig. \ref{fig:concept_tree}, each support candidate brings forth $L$ child candidates in the MMP algorithm.
In the $k$-th iteration, each support candidate chooses indices of $L$ columns that are maximally correlated with the residual. Each of chosen indices, in conjunction with previously selected indices, constructs a new support candidate.
For each support candidates, an estimate of the desired signal and also the residual for the next iteration are generated.
Specifically, let $\hat{\mathcal{E}}^{k-1}_j = \{ t_1, \cdots, t_{k-1} \}$ be the $j$-th support candidate in the $(k-1)$-th iteration, then the set of $L$ indices chosen from this support candidate, denoted as $\mathcal{E}^*$, is expressed as
$\mathcal{E}^* = \arg \mathop {\max} \limits_{ \{ \mathcal{E}: \left | \mathcal{E} \right | = L\}} { \| ( \mathbf{H}^H \mathbf{r}_j^{k-1} )_{\mathcal{E}} \|_2^2 }$
where $(\mathbf{\cdot})_{\omega}$ denotes construction of a vector from the support $\omega$. The residual $\mathbf{r}_j$ and the estimate $\mathbf{\hat e}_j$ of the desired signal are expressed as
\begin{eqnarray}
\mathbf{r}_j^{k-1} &=& \mathbf{y}' - \sqrt{P} \mathbf{H}_{\hat{\mathcal{E}}^{k-1}_j }\mathbf{\hat e}_j \\
\mathbf{\hat e}_j &=& \frac{1}{\sqrt{P}}(\mathbf{H}_{\hat{\mathcal{E}}^{k-1}_j }^T \mathbf{H}_{ \hat{\mathcal{E}}^{k-1}_j })^{-1} \mathbf{H}_{\hat{\mathcal{E}}^{k-1}_j }^T \mathbf{y}' . \nonumber 
\end{eqnarray}
The newly generated support candidates, which are the child candidates of $\hat{\mathcal{E}}^{k-1}_j$, are expressed as $\hat{\mathcal{E}}^{k-1}_j \bigcup \{ \mathcal{E}^* (i) \}$ for $i = 1, \cdots, L$.
Since multiple promising support candidates are investigated, it is not hard to convince oneself that the performance of the MMP algorithm is better than the sequential greedy algorithm returning a single support candidate. In fact, it has been shown that the MMP performs close to the best possible estimator using genie support information (called Oracle estimator) for high SNR regime \cite[Theorem 4.6]{mmp}. The operation of MMP is summarized in Table \ref{tab:mmp_alg}.


\section{Performance Analysis}
\label{sec:analysis}
%
In this section, we analyze the performance of the proposed PSED algorithm.
%
%
%
%
As a measure of performance, we consider the normalized MSE between the original symbol vector $\mathbf{s}$ and the output of PSED algorithm $\doublehat{\mathbf{s}}$, which is defined as
\begin{eqnarray}
{\rm MSE} &=& \frac{1}{n_t}E [ \| \mathbf{s} - \doublehat{\mathbf{s}} \|^2] \\
          &=& \frac{1}{n_t}E [\| \mathbf{s} - (\mathbf{s} + \hat{\mathbf{e}} - \mathbf{e} ) \|^2] \\
          &=& \frac{1}{n_t}E [\|\mathbf{e} - \hat{\mathbf{e}}\|^2]
\label{eq:mse2}
\end{eqnarray}
One can observe that the MSE associated with the symbol vector $\mathbf{s}$ is equivalent to the MSE associated with the error vector $\mathbf{e}$.
Our analysis is asymptotic in nature (i.e., dimension of the channel matrix is very large) since we use random matrix theory for analytical tractability.
In our analysis, we assume that the SNR  is high enough  so that the error vector $\mathbf{e}$ produced by the conventional detector can be modeled as a $K$-sparse vector (i.e., the number of nonzero entries is no more than $K$).
%
We first inspect the capability of the MMP algorithm to find the true support of $\mathbf{e}$ for random matrix $\mathbf{H}$. To this end, we define the restricted isometry property (RIP) of the channel matrix $\mathbf{H}$ as
%
%
%
\begin{definition}[RIP \cite{rip}]
A matrix $\mathbf{H} \in {\mathcal R}^{n_r \times n_t}$ is said to meet the RIP condition if the following inequality is satisfied for all $K$-sparse signals;
\begin{align}
(1-\delta_{K})\|\mathbf{x}\|_2^{2} \leq \|\mathbf{H}\mathbf{x} \|_2^2 \leq (1+\delta_{K})\|\mathbf{x}\|_2^{2}.
\end{align}
The smallest  $\delta_{K}$ satisfying the above RIP condition is called RIP constant.
\end{definition}
This RIP condition has been popularly used to identify the performance guarantees for many sparse recovery algorithms. For example, the exact recovery condition for the BP in noiseless condition is given by $\delta_{2K}< \sqrt{2}-1$.
Following theorem describes the recovery guarantee of MMP for the noiseless scenario.
\begin{theorem}[Exact recovery condition of MMP \cite{mmp}]
\label{theorem:mmp}
The MMP recovers a $K$-sparse signal $\mathbf{x}$ from the noiseless measurements $\mathbf{y} = \mathbf{H}\mathbf{x}$ accurately if $\mathbf{H}$ satisfies
\begin{align} \label{eq:mmpth}
\delta_{K+L} < \frac{\sqrt{L}}{\sqrt{K}+2\sqrt{L}},
\end{align}
where $L$ is the number of the child support candidates chosen from each parent candidate.
\end{theorem}
Note that when $K=L$, the recovery condition of the MMP becomes $\delta_{2K} <1/3$, which is close to the condition of the BP. When a signal is perturbed by noise vector $\mathbf{v}$, the exact recovery is not possible, and our interest lies in the condition under which the signal support is identified accurately.
\begin{theorem}[Exact support recovery condition for MMP \cite{mmp}]
\label{theorem:mmp2}
Let $e_i$ be the $i$th element of the signal vector $\mathbf{e}$.
If (\ref{eq:mmpth}) is satisfied and
\begin{align} \label{eq:supr}
\min_{i \in [1,2,\cdots, N]} |e_i| \geq  \tau \|\mathbf{v}\|_2,
\end{align}
where $\tau= \max(\gamma,\mu,\lambda)$, $\gamma=\frac{\sqrt{1+\delta_{L+K}}(\sqrt{L}+\sqrt{K})}{\sqrt{LK}-\left(\sqrt{LK}+K\right)\delta_{L+K}}$, $\mu = \frac{\sqrt{1+\delta_{L+K}}(1-\delta_{L+K})(\sqrt{L}+\sqrt{K})}{\sqrt{L}-(2\sqrt{L}+\sqrt{K})\delta_{L+K}}$, and
$\lambda = \sqrt{\frac{2(1-\delta_K)^2}{(1-\delta_K)^3 - (1+\delta_K)\delta_{2K}^{2}}}$, then the MMP recovers the true support.
\end{theorem}
Note that the condition in (\ref{eq:supr}) implies that the smallest magnitude of the signal entries should exceed the $\ell_2$-norm of the noise vector by a constant depending on the RIP constant.

Now we show that this support recovery conditions are satisfied with high probability for certain class of random matrices.
\begin{theorem}[RIP of Gaussian random matrix \cite{baraniuk_rip}]
\label{theorem:random}
If the elements of the random matrix $\mathbf{H}$ have i.i.d. Gaussian entries with ${\mathcal N}(0,1/\sqrt{m})$,
then there exist constants $c_1, c_2 >0$ depending only on $\delta$ such that the RIP holds with the prescribed $\delta$ and any $n_r \geq c_1K \log(n_t/K)$ with probability $\geq 1 - 2 e^{-c_2 n_r}$.
\end{theorem}
Theorem \ref{theorem:random} states that if the number of measurements is sufficiently large, i.e., $n_r \geq c_1K \log(n_t/K)$, Gaussian random matrices satisfy the RIP condition with overwhelming probability. Using this together with Theorem 3, we conclude that if the $\ell_2$-norm of the noise vector is sufficiently small, the condition in (\ref{eq:supr}) is satisfied with high probability.
Indeed, since the left-hand term in (\ref{eq:supr}) is determined by the minimum distance $d$ between adjacent symbol constellation points, we can check that the condition in (\ref{eq:supr}) is satisfied for high SNR regime.
If we assume that noise vector is Gaussian distributed, i.e., $\mathbf{v} \sim \mathcal{CN}(0, \sigma_v^2\mathbf{I})$, $\|\mathbf{v}\|_2^2$ has Chi-square distribution with degree of freedom $2n_r$ \cite{papoulis}. Hence, the probability that the condition (\ref{eq:supr}) is met is expressed in terms of the CDF of $\|\mathbf{v}\|_2^2$, i.e.,
\begin{align} \label{eq:cdf}
Pr\left(  \|\mathbf{v}\|_2^2 \leq \frac{d^2}{\tau^2} \right)  =  1-\frac{\Gamma(n_r,\frac{d^2}{\sigma_v^2\tau^2})}{\Gamma(n_r)}
\end{align}
where $\Gamma(s,x)=\int_{x}^{\infty} t^{s-1} e^{-t} dt$ and $\Gamma(s)=\Gamma(s,0)$.
Note that due to the asymptotic behavior $\Gamma(s,x) \sim x^{s-1}e^{-x}$ as $x \rightarrow \infty$, the term $\Gamma(n_r,\frac{d^2}{\sigma_v^2\tau^2})$ decays exponentially to zero as $1/\sigma_v^2 \rightarrow \infty $.
In high SNR regime, therefore, we observe that the probability in (\ref{eq:cdf}) approaches one. In what follows, we assume that the MMP identifies the support of $\mathbf{e}$ accurately.

\begin{figure*} [t]
 \centering
   {\includegraphics[width=110mm,height=85mm]{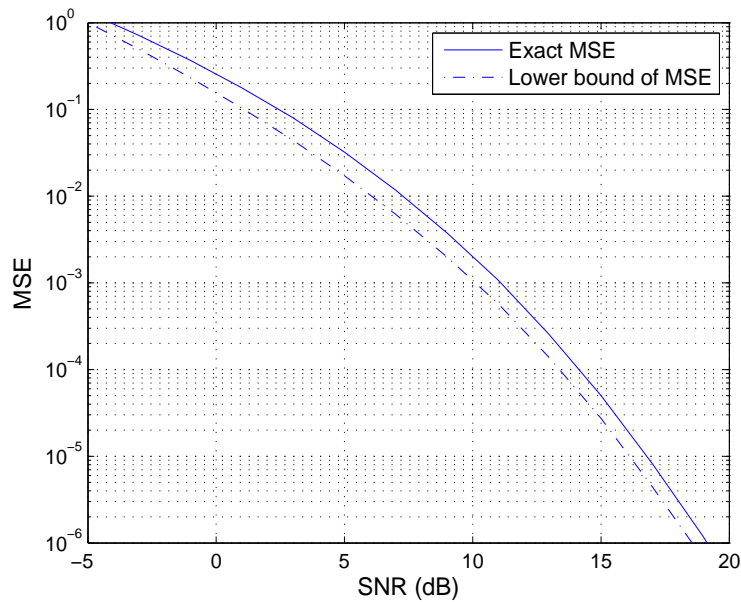}}
   \caption {Comparison of the exact MSE in \eqref{eq:beforeconj} and lower bound in the right-hand side of \eqref{eq:mse_lb0} for i.i.d. complex Gaussian random matrix. We set $n_r=n_t=128$ and use BPSK modulation.
   } \label{fig:emp}
\end{figure*}

When the support of $\mathbf{e}$ is identified, all non-support elements in $\mathbf{e}$ and columns of $\mathbf{H}$ associated with these can be removed from the system model. The resulting overdetermined system model is\footnote{$\mathbf{H}_D$ is a submatrix of $\mathbf{H}$ that only contains columns indexed by $D$. For example, if $D = \{1, 4, 5\}$, then $\mathbf{H}_D = [\mathbf{h}_1 ~ \mathbf{h}_4 ~ \mathbf{h}_5]$ where $\mathbf{h}_j$ is the $j$-th column of $\mathbf{H}$.}
\begin{align}
\mathbf{y}' &= \sqrt{P} \mathbf{H} \mathbf{e} + \mathbf{v}\\
&= \sqrt{P} \mathbf{H}_\mathcal{E} \mathbf{e}_\mathcal{E} + \mathbf{v}.
\label{eq:mse4}
\end{align}
%
Note that most of greedy sparse recovery algorithms use the linear squares (LS) solution in generating the estimate of error vector $\hat{\mathbf{e}}_\mathcal{E}$. In this case, the estimate $\hat{\mathbf{e}}_\mathcal{E}$ is given by \cite{kay}
%
\begin{equation} \label{eq:zff}
\hat{\mathbf{e}}_\mathcal{E} =\frac{1}{\sqrt{P}}{\left(\mathbf{H}_{\mathcal{E}}^{H}\mathbf{H}_{\mathcal{E}}\right)^{-1}\mathbf{H}_{\mathcal{E}}^{H}}\mathbf{y}'.
\end{equation}
%
%
%
%
From (\ref{eq:mse4}) and (\ref{eq:zff}), the MSE is expressed as
\begin{align}
{\rm MSE}_{psed} &=  \frac{1}{n_t} E [ \| \mathbf{e} - \hat{\mathbf{e}} \|^2] \\
&= \frac{1}{n_t} E [\|\mathbf{e}_\mathcal{E} - \hat{\mathbf{e}}_\mathcal{E} \|^2] \\
&=  E_{\mathbf{H}} \left[ {\rm tr}E_{\mathcal{E},\mathbf{v}}\left[\frac{1}{n_t P}\left(\mathbf{H}_{\mathcal{E}}^{H}\mathbf{H}_{\mathcal{E}}\right)^{-1}\mathbf{H}_{\mathcal{E}}^{H}\mathbf{v}\mathbf{v}^{H} \mathbf{H}_{\mathcal{E}} \left(\mathbf{H}_{\mathcal{E}}^{H}\mathbf{H}_{\mathcal{E}} \right)^{-1}\big|\mathbf{H} \right] \right], \label{eq:beforeconj}
\end{align}
where $E_{\mathbf{x}}[\cdot]$ denotes the expectation with respect to the random variable $\mathbf{x}$.
One thing to notice is that $\mathbf{H}_{\mathcal{E}}$ and $\mathbf{v}$ are correlated with each other since the error pattern associated with $\mathbf{H}_{\mathcal{E}}$ depends on the realization of the noise vector $\mathbf{v}$. Since this makes the evaluation of (\ref{eq:beforeconj}) very difficult,
we take an alternative approach and investigate the lower bound of the MSE.
First, let $\tilde{\mathcal{E}}$ be the set of indices whose elements are randomly chosen from the set of all column indices $\Omega = \{ 1, 2, \cdots, n \}$ and the cardinality of $\tilde{\mathcal{E}}$ is the same as that of $\mathcal{E}$ (i.e., $| \tilde{\mathcal{E}} | = | \mathcal{E} |$).
Then, we conjecture that
\begin{align}
{\rm MSE}_{psed}
 & \geq  E_{\mathbf{H}} \left[  {\rm tr}E_{\mathcal{E},\mathbf{v}}\left[\frac{1}{n_t P}\left(\mathbf{H}_{\tilde{\mathcal{E}}}^{H}\mathbf{H}_{\tilde{\mathcal{E}}}\right)^{-1}\mathbf{H}_{\tilde{\mathcal{E}}}^{H}\mathbf{v}\mathbf{v}^{H} \mathbf{H}_{\tilde{\mathcal{E}}} \left(\mathbf{H}_{\tilde{\mathcal{E}}}^{H}\mathbf{H}_{\tilde{\mathcal{E}}}\right)^{-1}\Big|\mathbf{H} \right] \right] \label{eq:mse_lb0} \\
 &= E_{\mathbf{H}} \left[\frac{1}{\rm SNR}  {\rm tr}E_{\mathcal{E},\mathbf{v}}\left[ \frac{1}{n_t} \left(\mathbf{H}_{\tilde{\mathcal{E}}}^{H}\mathbf{H}_{\tilde{\mathcal{E}}}\right)^{-1}\Big|\mathbf{H}\right] \right]. \label{eq:mse_lb}
\end{align}
where ${\rm SNR} = \frac{P}{\sigma_v^2}$.
Note that this conjecture is justified by the fact that columns associated with $\mathcal{E}$
are caused by the detection errors while those associated with $\tilde{\mathcal{E}}$ are randomly chosen so that the former would yield higher MSE than the latter.
To judge the effectiveness of the conjecture, we performed the empirical simulations of two quantities (i.e., \eqref{eq:beforeconj} and the right-hand side of \eqref{eq:mse_lb0}) for i.i.d. complex Gaussian random matrix.
We observe from Fig. \ref{fig:emp} that the conjecture holds true empirically and also the lower bound in \eqref{eq:mse_lb0} is tight across the board.

We next investigate an asymptotic behavior of the term $E\left[ \frac{1}{\rm SNR}  {\rm tr}E\left[ \frac{1}{n_t} \left(\mathbf{H}_{\tilde{\mathcal{E}}}^{H}\mathbf{H}_{\tilde{\mathcal{E}}}\right)^{-1}\Big|\mathbf{H}\right] \right]$.
We assume that the elements of the channel matrix $\mathbf{H}$ are i.i.d. complex Gaussian ($h_{ij} \sim \mathcal{CN}(0,1/n_r)$) and the dimension of the channel matrix $\mathbf{H}$ is sufficiently large (i.e., $n_t, n_r \rightarrow \infty$) with a constant ratio $\frac{n_t}{n_r} \rightarrow \beta$.
Let $\mathbf{H}_{\tilde{\mathcal{E}}}$ be the submatrix generated by randomly choosing $| \mathcal{E} |$ columns of $\mathbf{H}$. Then we have
$$\frac{1}{| \mathcal{E} |} {\rm tr} \left(\mathbf{H}_{\tilde{\mathcal{E}}}^{H}\mathbf{H}_{\tilde{\mathcal{E}}}\right)^{-1} \rightarrow \frac{1}{1-\beta'}$$
where $\beta'=\frac{| \mathcal{E} |}{n_r}$ \cite[Section 2.3.3]{tulino}.
Notice that when the dimension of the matrix is large, the quantity converges to the deterministic limit depending on the ratio $\beta'$ irrespective of channel realization.
Therefore, it suffices to evaluate a single matrix realization in \eqref{eq:mse_lb} and hence the outer expectation (with respect to the channel realization) is unnecessary.

%
%
%
We next investigate the distribution of $|\mathcal{E}|$ (cardinality of $\mathcal{E}$) which corresponds to the number of errors in the output streams of the linear detector. Here we consider the LMMSE detector as a conventional linear detector.
Note that the event of detection errors is related to the post-detection signal-to-interference-and-noise-ratio (SINR) at the output of the LMMSE detector.
While the SINR for each output stream relies on a channel realization, when the system size gets large (i.e., $n_t, n_r \rightarrow \infty$ and $\frac{n_t}{n_r} \rightarrow \beta$), the SINR for all output streams approaches
the same quantity
\cite{tulino}
\begin{align}
\mathrm{SINR}_i \longrightarrow \mathrm{SINR}_{\infty} = {\rm SNR} - \frac{\mathcal{F}({\rm SNR},\beta)}{4} \nonumber
\end{align}
where $\mathcal{F}(x,z) = \left(\sqrt{x(1+\sqrt{z})^{2}+1} - \sqrt{x(1-\sqrt{z})^{2}+1}\right)^{2}$.

In addition, one can show that the output streams of the LMMSE receiver are asymptotically uncorrelated with each other  (see Appendix \ref{apx:uncor}).
Thus, the detection problem for each output stream can be considered separately and the number of errors $|\mathcal{E}|$ is approximated as a Binomial distribution with the success probability $P_{e}$, where $P_e$ is the probability of error event for each output stream. That is,
$$Pr(| \mathcal{E} |=t) \approx \left(\begin{matrix}n_t\\t\end{matrix}\right)P_{e}^t(1-P_{e})^{n_t-t}$$
When $n_t$ is large, the Binomial distribution with parameter $n_t$ and $P_e$ approaches to the Normal distribution $\mathcal{N}(n_t P_{e}, ~ n_t P_{e}(1-P_{e}))$ by DeMoivre-Laplace theorem \cite{stark} and hence one can show that (see Appendix \ref{apx:conv})
\begin{align}
&P_r \left(\left|\frac{| \mathcal{E} |}{n_r} - P_{e}\beta \right|>\epsilon \right) \rightarrow 0  \label{eq:dl1} \\
&P_r \left(\left|\frac{| \mathcal{E} |}{n_t} - P_{e}\right|>\epsilon \right) \rightarrow 0.
 \label{eq:dl2}
\end{align}
%
%
 %
 %
Our discussions so far can be summarized as follows.
\begin{enumerate}
\item $\frac{1}{| \mathcal{E} |} {\rm tr} \left(\mathbf{H}_{\tilde{\mathcal{E}}}^{H}\mathbf{H}_{\tilde{\mathcal{E}}}\right)^{-1}$  converges to $\frac{1}{1 - \beta'} =\frac{1}{1- \frac{ | \mathcal{E} | }{n_r } }$ irrespective of the channel realization.
    %
    %
\item $\frac{| \mathcal{E} |}{n_r}$ and $\frac{| \mathcal{E} |}{n_t}$ converge in probability to $P_{e}\beta$ and $P_{e}$, respectively.
\end{enumerate}
Using these, one can show that
\begin{eqnarray}
\frac{1}{\rm SNR}  {\rm tr}E\left[ \frac{1}{n_t} \left(\mathbf{H}_{\tilde{\mathcal{E}}}^{H}\mathbf{H}_{\tilde{\mathcal{\mathcal{E}}}}\right)^{-1}\Big|\mathbf{H}\right]
&=& \frac{1}{\rm SNR} E\left[ \frac{|\mathcal{E}|}{n_t} \frac{1}{ |\mathcal{E}|}{\rm tr} \left(\mathbf{H}_{\tilde{\mathcal{E}}}^{H}\mathbf{H}_{\tilde{\mathcal{\mathcal{E}}}}\right)^{-1}\Big|\mathbf{H}\right] \label{eq:uuu1} \\
  &\rightarrow &  \frac{1}{\rm SNR} E\left[ \frac{ |\mathcal{E} |}{n_t } \frac{ 1 }{1- \frac{ |\mathcal{E}| }{n_r }}\right]  \label{eq:uuu2} \\
   &=& \frac{1}{\rm SNR} \frac{n_r}{n_t} E\left[  \frac{ \frac{ |\mathcal{E} |}{n_r } }{1- \frac{ |\mathcal{E}| }{n_r }}\right] \label{eq:uuu3} \\
   &\geq& \frac{1}{\rm SNR} \frac{n_r}{n_t}  \frac{ E\frac{ |\mathcal{E} |}{n_r } }{1- E \frac{ |\mathcal{E}| }{n_r }} \label{eq:uuu4} \\
   &= & \frac{1}{\rm SNR} \frac{1}{\beta} \frac{ P_e \beta }{1- P_e \beta }  \label{eq:uuu5} \\
  &=& \frac{1}{\rm SNR} \frac{P_{e} }{1-P_{e}\beta}. \label{eq:uuu6}
 \end{eqnarray}
where \eqref{eq:uuu4} is from Jensen's inequality.
Noting that $\frac{ |\mathcal{E} |}{n_r } \ll 1$, we see that the obtained lower bound is tight.\footnote{$f(x) = \frac{x}{1-x}$ is a convex function for $0 < x < 1$ and hence $E \left[ \frac{x}{1-x} \right] \geq \frac{E[x]}{1-E[x]}$. In our case, $x = \frac{ |\mathcal{E} |}{n_r }$ and hence $x \ll 1$ so that $f(x) \approx x$ and the lower bound is tight.}

In the high SNR regime, $P_e \ll 1$ and hence
\begin{eqnarray}
{\rm MSE}_{psed} > \frac{1}{\rm SNR} {\rm tr}E\left[ \frac{1}{n_t} \left(\mathbf{H}_{\tilde{\mathcal{E}}}^{H}\mathbf{H}_{\tilde{\mathcal{\mathcal{E}}}}\right)^{-1}\Big|\mathbf{H}\right] \gtrapprox \frac{P_{e}}{\rm SNR}. \label{eq:simple_form}
 \end{eqnarray}
Since $P_e$ is a function of SINR, the obtained lower bound of ${\rm MSE}_{psed}$ is a function of SNR and SINR.
Indeed, when $n_t, n_r \rightarrow \infty$ and $\frac{n_t}{n_r} \rightarrow \beta$, the residual interference plus noise for the LMMSE detector approaches to Normal distribution \cite{zhang,guo} so that $P_{e}$ can be readily expressed as a function of the SINR.
For example, for the binary phase shift keying (BPSK), the error probability $P_e$ in symbol detection approaches  \cite{proakis} $$P_e = {\rm Erf}(\mathrm{SINR}_{\infty}) = Q \left( \sqrt{ 2\mathrm{SINR}_{\infty}} \right)$$
where $Q(x) = \int_{x}^{\infty} \frac{1}{\sqrt{2\pi }} \exp (-\frac{t}{2}) dt$.
Using $Q(x) > \frac{x}{1+x^2}\frac{1}{\sqrt{2\pi}}e^{-x^2/2}$, we have
\begin{align}
{\rm MSE}_{psed} &> \frac{Q \left( \sqrt{ 2\mathrm{SINR}_{\infty}} \right)}{\rm SNR} \nonumber  \\
&> \frac{1}{\rm SNR} \frac{\sqrt{2\mathrm{SINR}_{\infty}}}{1+2\mathrm{SINR}_{\infty}}\frac{1}{\sqrt{2\pi}} \exp\left(-\mathrm{SINR}_{\infty}\right)
\end{align}
In high SNR regime, we have \cite{verdu}
\begin{align}
\mathrm{SINR}_{\infty} \approx \left\{ \begin{array}{cc} \sqrt{{\rm SNR}}&  \beta = 1 \\ (1-\beta){\rm SNR}+\frac{\beta}{1-\beta} & \beta < 1  \end{array} \right.
\end{align}
and hence
\begin{eqnarray}
{\rm MSE}_{psed} \gtrapprox \left\{ \begin{array}{cc} \frac{1}{2\sqrt{\pi}} \frac{1}{{\rm SNR}^{5/4}} \exp\left(-\sqrt{{\rm SNR}}\right)  & \text{if } \beta = 1, \\ \frac{1}{2\sqrt{\pi(1-\beta)}} \frac{1}{{\rm SNR}^{3/2}} \exp\left(-(1-\beta){\rm SNR}\right)  & \text{if } \beta < 1 \end{array} \right..
  \label{eq:nmse_prop}
\end{eqnarray}

On the other hand, ${\rm MSE}$ of the conventional linear MMSE detector is given by \cite{tulino}
\begin{align}
{\rm MSE}_{conv} &= \frac{1}{n_t} E_{\mathbf{H}} \left[ {\rm  tr}(\mathbf{I} + {\rm SNR} \mathbf{H}^{H} \mathbf{H})^{-1} \right] \nonumber \\
& \longrightarrow  1 - \frac{\mathcal{F} ({\rm SNR}, \beta)}{4 \beta {\rm SNR}}.
                \label{eq:mse_conv}
\end{align}
Further, in high SNR regime, ${\rm MSE}_{conv}$ can be expressed as\cite[Chap 6.3]{verdu}
\begin{eqnarray}
{\rm MSE}_{conv} &\rightarrow&  1 - \frac{\mathcal{F} ({\rm SNR}, \beta)}{4 \beta {\rm SNR}} \nonumber  \\
 &\approx& \left\{
\begin{array}{rl}
\frac{1}{\sqrt{{\rm SNR}}}  & \text{if } \beta = 1, \\
\frac{1}{(1 - \beta) {\rm SNR}}  & \text{if } \beta < 1.
\end{array} \right.
  \label{eq:nmse_conv}
\end{eqnarray}
%
%

\begin{figure*} [t]
 \centering
  \subfigure[]
  {\includegraphics[width=85mm,height=70mm]{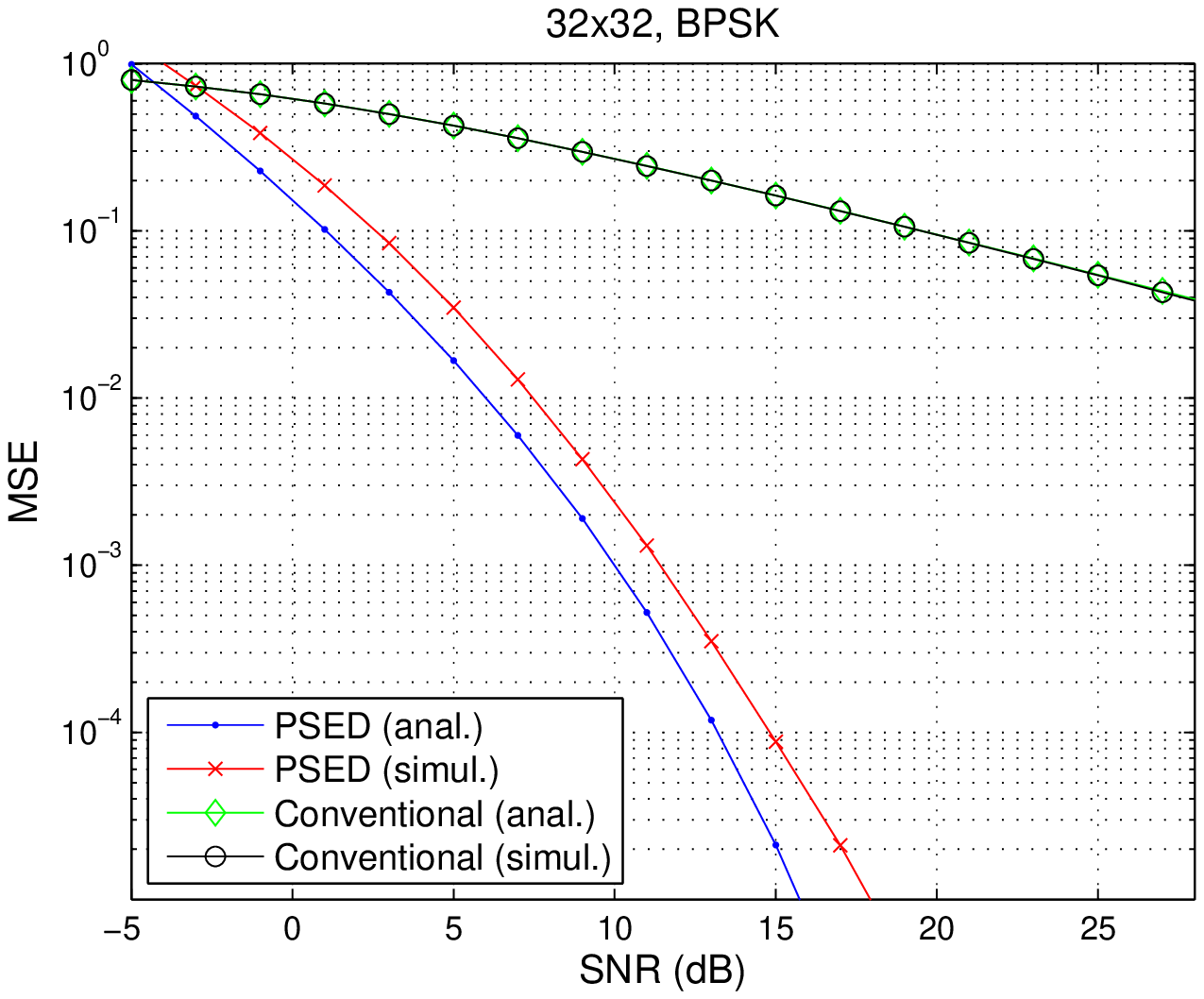}}
   \hspace{-1cm}
  \subfigure[]
   {\includegraphics[width=85mm,height=70mm]{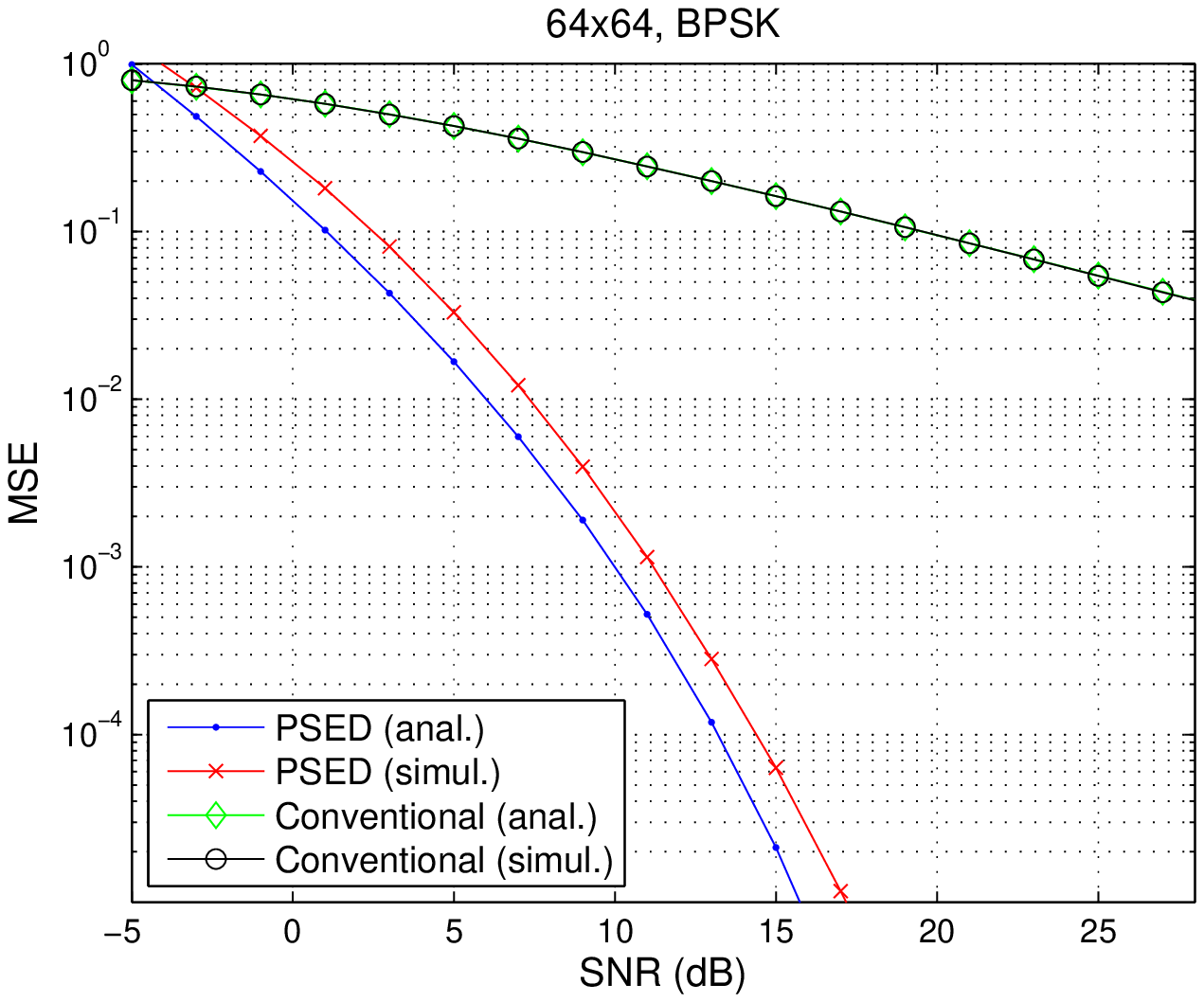}}
     \subfigure[]
  {\includegraphics[width=85mm,height=70mm]{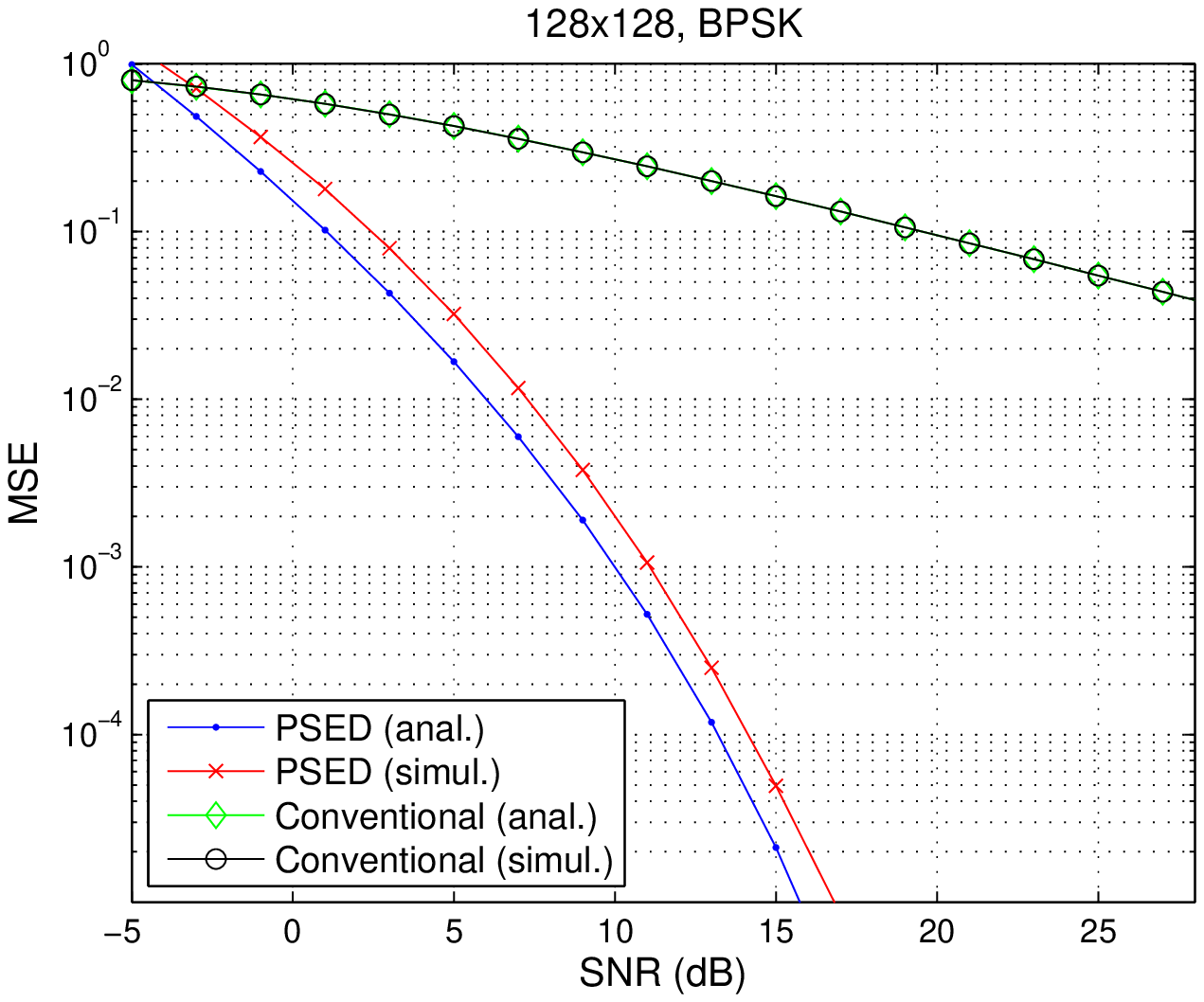}}
   \caption {The MSE performance of the PSED technique and the conventional detector: (a) $n_r=n_t=32$, (b) $n_r=n_t=64$, and (c) $n_r=n_t=128$.
   BPSK modulation is used.
   } \label{fig:empirical}
\end{figure*}

%
It is interesting to compare \eqref{eq:nmse_prop} and \eqref{eq:nmse_conv}, asymptotic MSEs for BPSK signals when the dimension of the matrix goes to infinity. While the MSE of the conventional method decays linearly or sublinearly with SNR, the MSE of the PSED decays exponentially with SNR.
In Fig. \ref{fig:empirical}, we plot the MSE in \eqref{eq:nmse_prop} and \eqref{eq:nmse_conv} and the empirical MSEs as a function of the SNR when i.i.d. complex Gaussian system matrix and BPSK modulation are used. In general, we observe that the obtained bound matches well with the simulation results. Noting that our analysis is asymptotic in nature, it is no wonder that the bound becomes more accurate when the dimension of the matrix increases.
We also observe that the performance difference between the PSED algorithm and the conventional detector is substantial and further the difference increases with SNR.
This is mainly because the quality of the estimated error vector $\hat{\mathbf{e}}$ for the sparse recovery algorithm improves with SNR so that the error difference $\epsilon = \hat{\mathbf{e}} - \mathbf{e}$ and the resulting MSE also get smaller.


\vspace{0.5cm}
\section{Computational Cost of Sparse Recovery} \label{sec:complexity}
%
%
In this subsection, we analyze the computational complexity of the PSED algorithm. We consider two versions of the PSED algorithm; PSED with LMMSE detection (PSED-LMMSE) and PSED with MF detection (PSED-MF).
As mentioned, additional operations caused by the proposed method are 1) sparse transform and 2) sparse error recovery.
While the computational overhead of the sparse transform is fixed (matrix multiplication and subtraction), that for the sparse error vector recovery depends on the tree branching parameter $L$ of MMP and the sparsity $K$. Since the number of iterations of MMP is set to the sparsity $K$, $K$ matrix inverse operations (from $1\times 1$ to $K\times K$ matrix inverse) are required.
Noting that the sparsity $K$ is much smaller than the dimension of symbol vector $n_t$ and the MMP algorithm performs well with small value of $L$ (in our simulations, we set $L = 2$), one can expect that the additional burden of the PSED would be marginal.

%
%
\begin{table*}[!]
\caption{The total number of complex multiplications of several detectors}
\begin{center}
\begin{tabular}{m{2.6cm}|m{1.5cm}|m{2.5cm}|m{2cm}|m{3.5cm}}
\hline
Operation & MF detector &  PSED-MF & LMMSE detector  &  PSED-LMMSE   \\ \hline\hline
Filter weight generation  & 0 & 0 & $2n_r n_t^{2} + I_v(n_t)$ & $2n_r n_t^{2} + I_v(n_t)$ \\ \hline
Filtering & $n_r n_t$ & $n_r n_t$  & $n_r n_t$ & $n_r n_t$  \\ \hline
Sparse transform & $0$    & $n_r n_t$ & 0     & $n_r n_t$   \\ \hline
Sparse recovery (matching) & $0$ & $\sum_{k=1}^{K} n_r (n_t-k+1)$ & $0$   & $\sum_{k=1}^{K} n_r (n_t-k+1)$  \\ \hline
Sparse recovery (orthogonal projection) & $0$ & $\sum_{k=1}^{K}(2n_rk^2+I_v(k)+k n_r)L$ & $0$ & $\sum_{k=1}^{K}(2n_rk^2+I_v(k)+k n_r)L$ \\ \hline
Sparse recovery (residual generation) & $0$ & $\sum_{k=1}^{K} kn_r L$ & $0$ & $\sum_{k=1}^{K} k n_r L $ \\ \hline
Total & $n_r n_t$ & $2n_r n_t + \sum_{k=1}^{K} n_r (n_t-k+1) + (2n_rk^2+I_v(k)+2k n_r)L$ &
$2n_r n_t^{2} + I_v(n_t) + n_r n_t$  & $2n_r n_t^{2} + I_v(n_t) + 2n_r n_t + \sum_{k=1}^{K} n_r (n_t-k+1) + (2n_rk^2+I_v(k)+2k n_r)L$ \\ \hline
\end{tabular}
\end{center}
\label{tb:comp}
\end{table*}

Table \ref{tb:comp} summarizes the number of complex multiplications required for PSED and the conventional detectors.
When the LMMSE detector is used, inversion of the covariance matrix is needed in the weight generation process. In Table \ref{tb:comp}, we denote the number of complex multiplications to perform the inversion of an $n \times n$ matrix by $I_v(n)$.
Since the required complexity to invert a matrix is cubic in the dimension $n$ of the matrix (i.e., $O(n^3)$), additional matrix inversion overhead (from $1\times 1$ to $K \times K$ dimensional matrix) of PSED-LMMSE is small when compared to the $n_t \times n_t$ dimensional covariance matrix inversion of LMMSE.\footnote{Note that gaussian elimination method for $n \times n$ inversion requires $(2n^3+3n^2-5n)/6$ multiplications \cite{fb}. See \cite{zimmermann} for more efficient implementation of matrix inversion.}
For example, the additional complexity of PSED-LMMSE for $n_t = n_r = 32$ is only $13\%$ and that for $n_t = n_r = 64$ is $20\%$.
%
When the MF is used, there is no inversion and thus the additional complexity associated with PSED might be relatively higher than that of PSED-LMMSE. In spite of the increased computational overhead, due to the relatively low-complexity operations, overall complexity of the PSED-MF would be much lower than the computational burden required for the LMMSE detector.

%

\vspace{1.0cm}
\section{Simulations and Discussions}	
\label{sec:simulation}

\subsection{Simulation Setup}
In this section, we examine the performance of the proposed PSED algorithm (PSED-LMMSE and PSED-MF).
For comparison, we test conventional linear receivers (MF and LMMSE detectors) and K-best detection algorithm \cite{kbest}. Also, as a performance lower bound, we consider the ML detector implemented via sphere decoding (SD) algorithm. Note that the K-best detector is one of the popularly used sub-optimal MIMO detectors and its complexity does not vary with channel realizations and SNR.
%
Note also that we only present the result of the ML detector for $32\times 32$ dimensional system since it is very hard to obtain the results of higher dimension due to computational complexity.

In our simulations, we construct the channel matrix $\mathbf{H}$ whose entries are chosen from i.i.d. complex Gaussian (i.e., $h_{i,j} \sim \mathcal{CN} (0, \frac{1}{n_r})$). In addition, we generate the transmit symbol vectors whose elements are chosen from quadrature phase shift keying (QPSK) constellation.
%
%
In order to check the worst-case performance, we always turn on the sparse recovery algorithm for the PSED algorithm  even in low SNR range where the error signal would not be sparse.
As mentioned, we use the MMP algorithm in recovering the sparse error vector.
The MMP parameter $L$ is fixed to $2$ and the sparsity parameter is set to $K = 0.15 n_t$. That is, we assume that $15\%$ of transmit symbols are in error after the conventional detection. Although this choice is a bit overestimated sparsity estimate, in particular for mid and high SNR regime, we observe that it induces marginal performance loss over the case using accurate value $K$.\footnote{While we do not optimize the sparsity parameter and use a fixed parameter, one can further optimize to achieve better performance using either the residual based stopping criterion or cross validation (CV). Refer to \cite{bshim_sparse,CrossVal} for details.}
Denoting the number of the symbols survived for each layer for the K-best detector by $m$, we set $m$ such that its complexity is comparable to that of the PSED-LMMSE (see the complexity comparison in Table \ref{tb:comp2}).
The SNR per each receive antenna is defined as
\begin{equation}
{\rm SNR}_{r} = \frac{n_t}{n_r}{\rm SNR}.
\end{equation}

\begin{table*}[!]
\caption{Comparison of computational complexity of several detectors}
\begin{center}
\begin{tabular}{m{2.5cm}|m{2.5cm}|m{2.5cm}|m{2.5cm}|m{2.5cm}}
\hline
& $32\times 32$ ($\times 10^3$) &  $64\times 64$ ($\times 10^3$) & $128 \times 128$ ($\times 10^3$) &  $256 \times 256$ ($\times 10^3$)   \\ \hline\hline
MF   & 1 & 4 & 16 & 66 \\ \hline
PSED-MF& 12 &135  &  1759 & 23439  \\ \hline
LMMSE& 78  & 618 & 4918 & 39245 \\ \hline
PSED-LMMSE& 88 & 744 & 6644 & 62553  \\ \hline
K-best  & 88 (m=15) & 778 (m=40) & 6666 (m=95) & 62305 (m=260)  \\ \hline
\end{tabular}
\end{center}
\label{tb:comp2}
\end{table*}

\subsection{Simulation Results} \label{sec:noisy_anal}


\begin{figure*}[t]
 \centering
 \subfigure[]
   {\includegraphics[width=85mm,height=75mm]{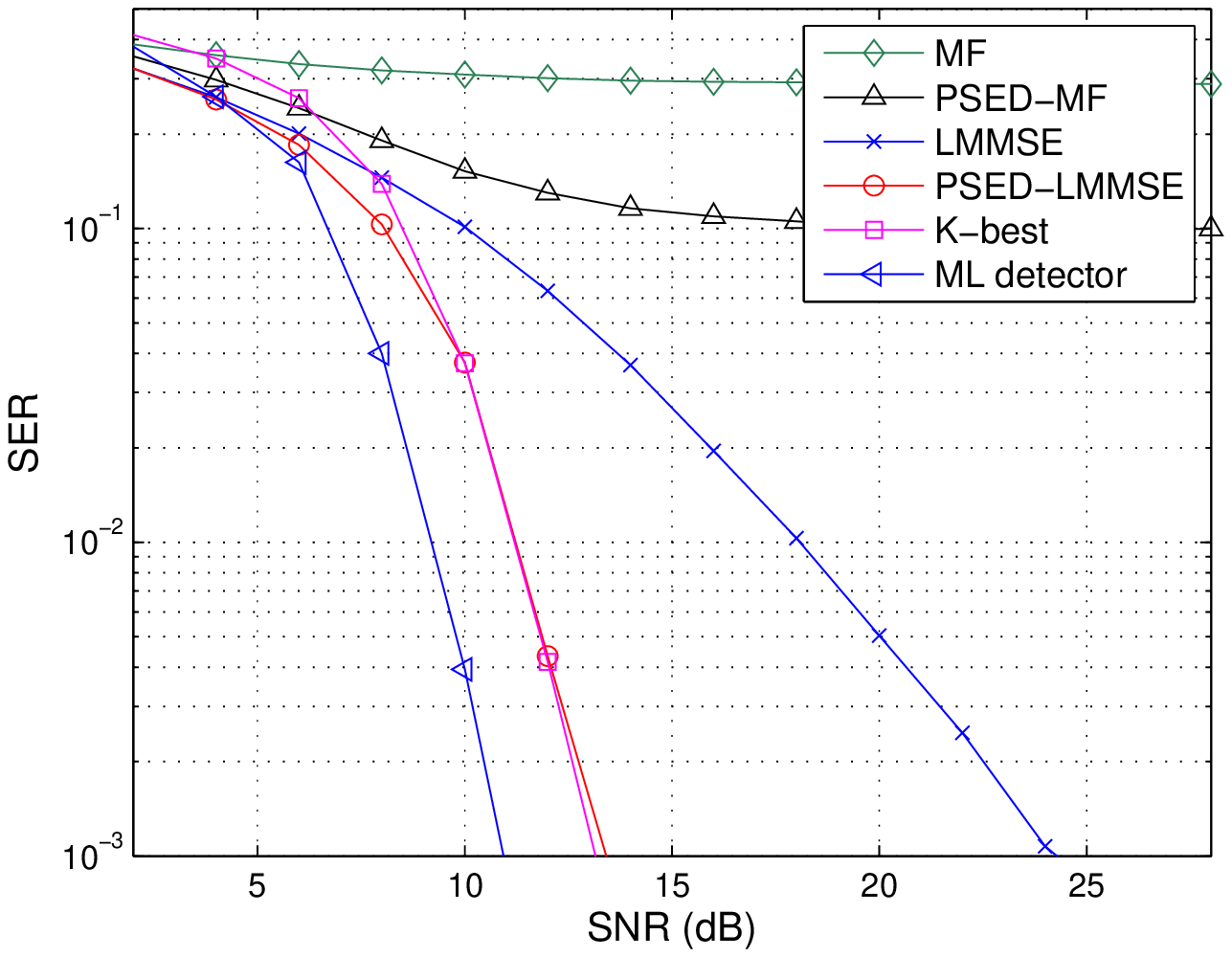}}
   \hspace{-1cm}
  \subfigure[]
   {\includegraphics[width=85mm,height=75mm]{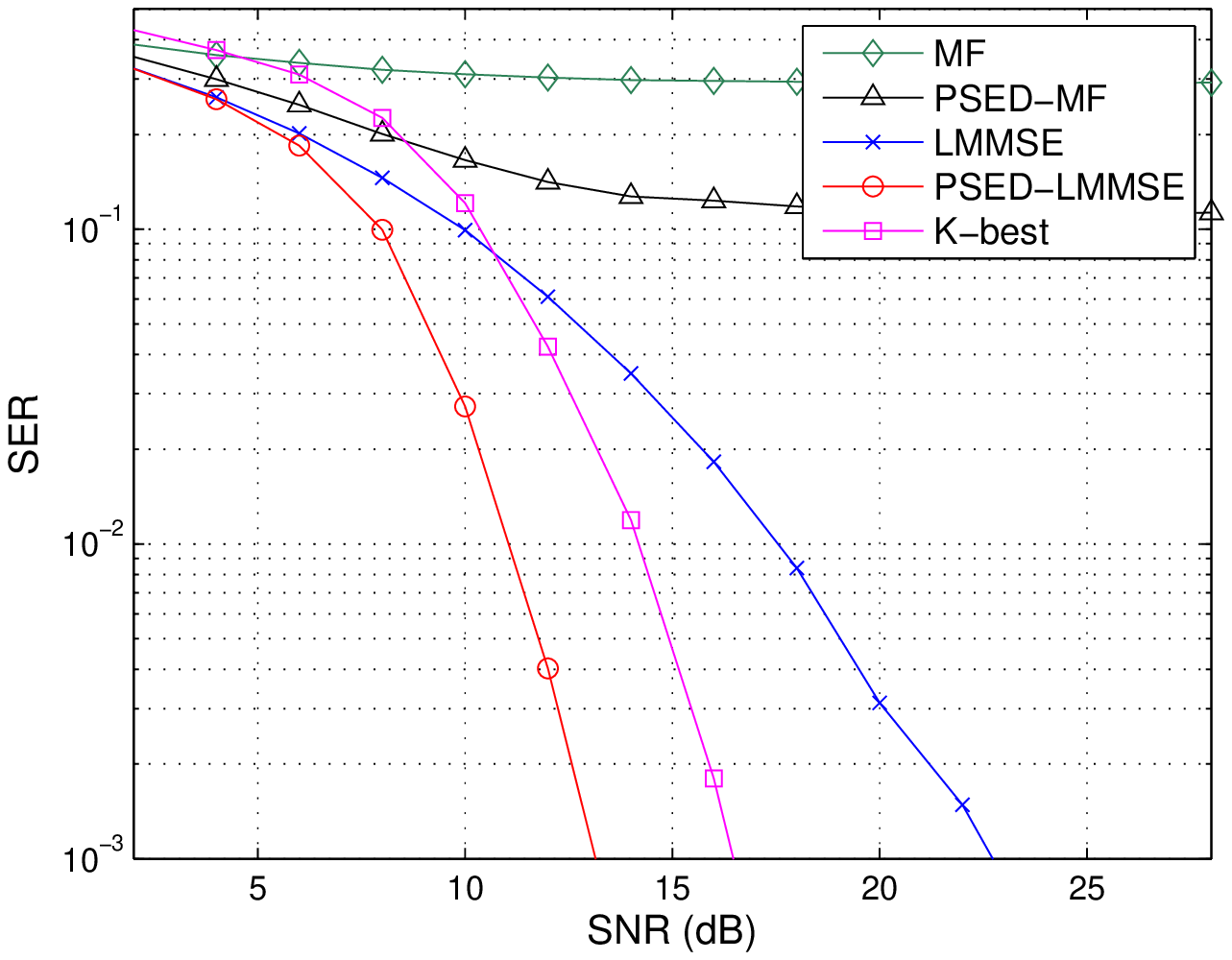}}
     \subfigure[]
  {\includegraphics[width=85mm,height=75mm]{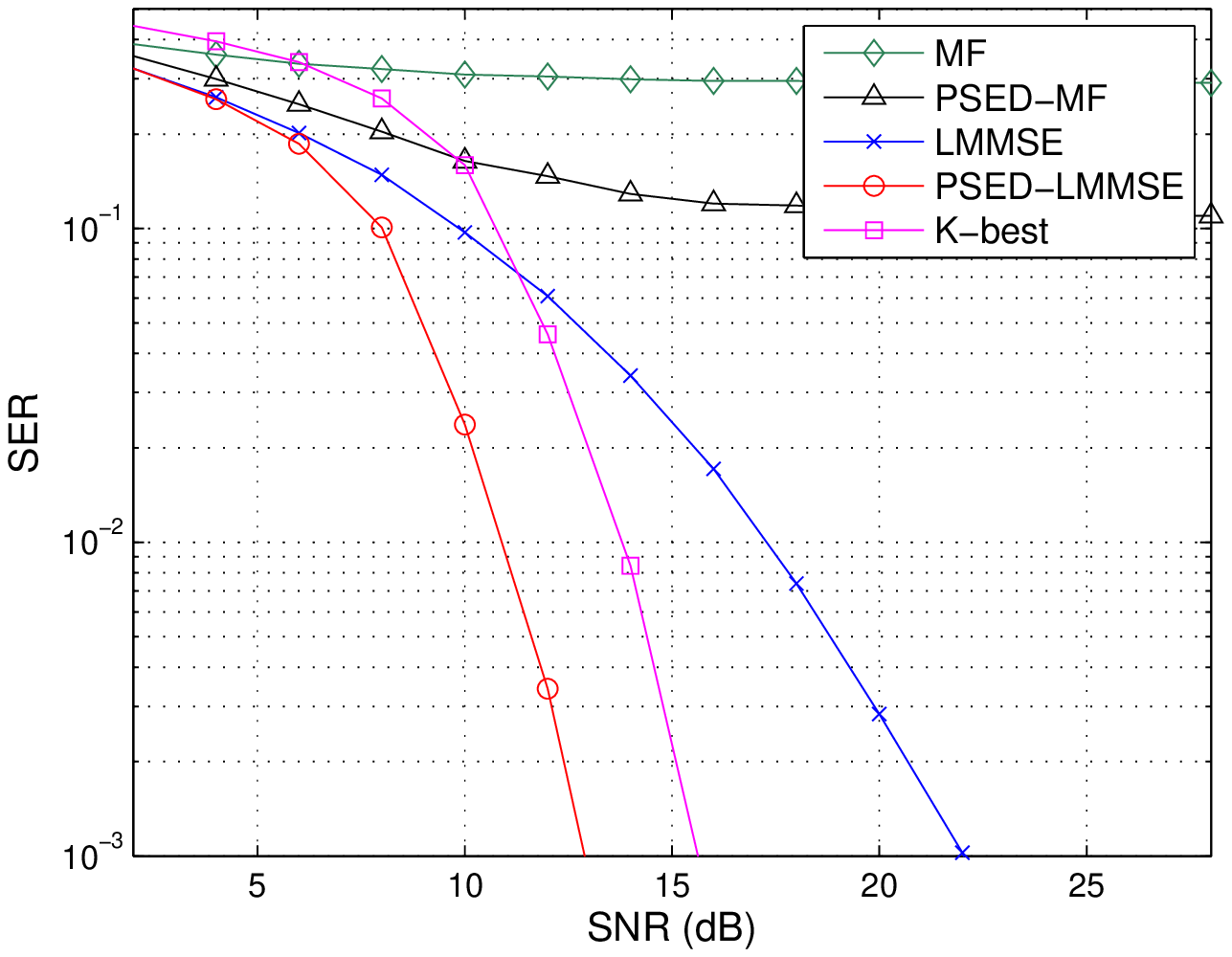}}
       \hspace{-1cm}
  \subfigure[]
   {\includegraphics[width=85mm,height=75mm]{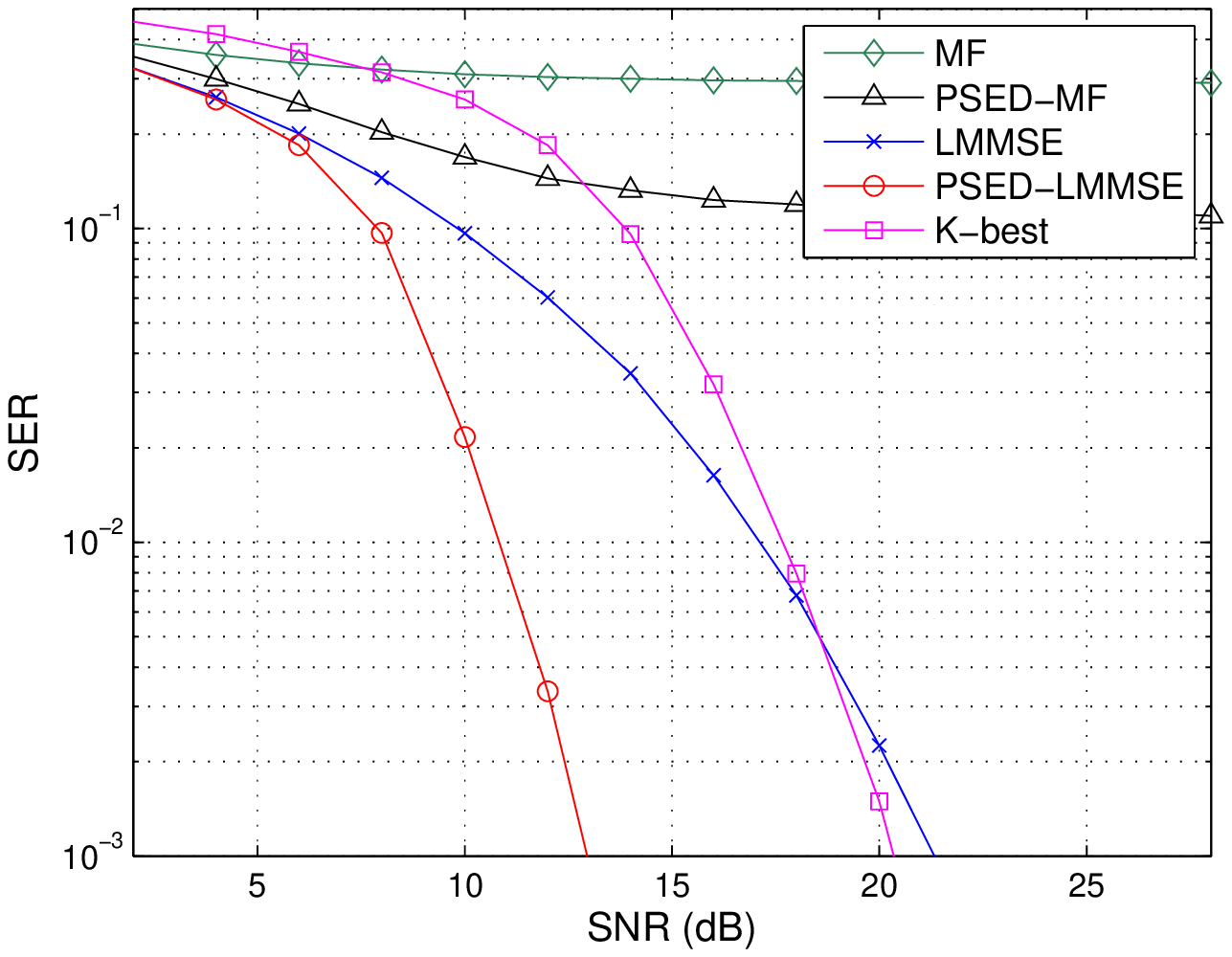}}
  \caption {The SER as a function of SNR for detectors. (a) $(n_r,n_t)=(32,32)$, (b) $(n_r,n_t)=(64,64)$, (c) $(n_r,n_t)=(128,128)$, and (d) $(n_r,n_t)=(256,256)$.
   } \label{fig:perf}
\end{figure*}

\begin{figure*} [t]
 \centering
 \subfigure[]
  {\includegraphics[width=85mm,height=75mm]{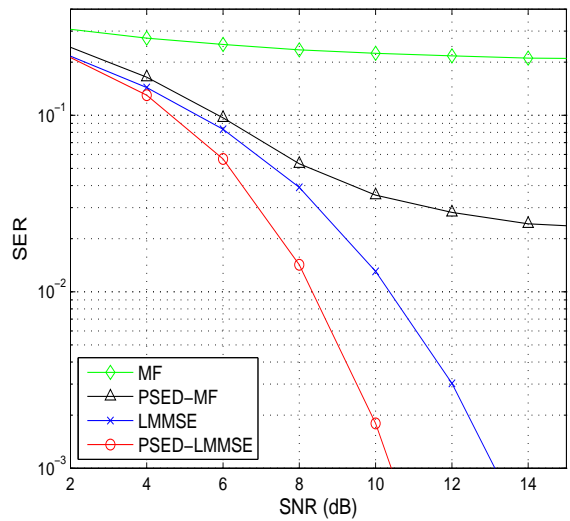}}
  \subfigure[]
  {\includegraphics[width=85mm,height=75mm]{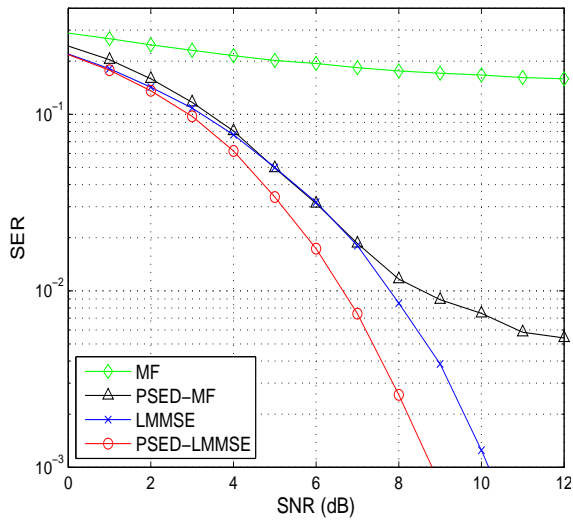}}
  \hspace{-1cm}
  \subfigure[]
   {\includegraphics[width=85mm,height=75mm]{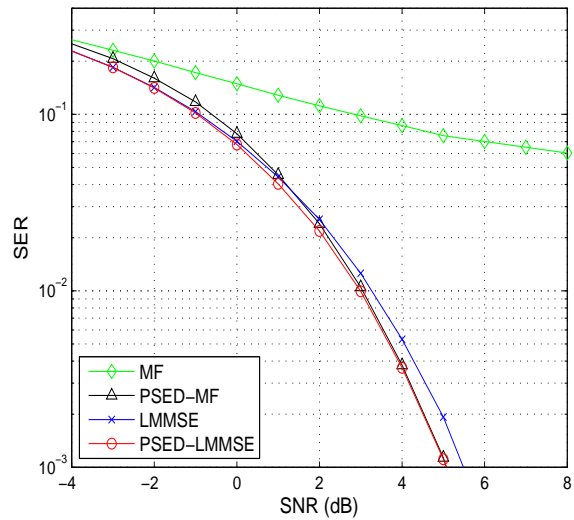}}

  \caption {The SER as a function of SNR for detectors. (a) $(n_r,n_t)=(48,32)$, (b) $(n_r,n_t)=(64,32)$, and (c) $(n_r,n_t)=(128,32)$.
   } \label{fig:perf2}
\end{figure*}

\begin{figure*}[t]
 \centering
 \subfigure[]
   {\includegraphics[width=85mm,height=75mm]{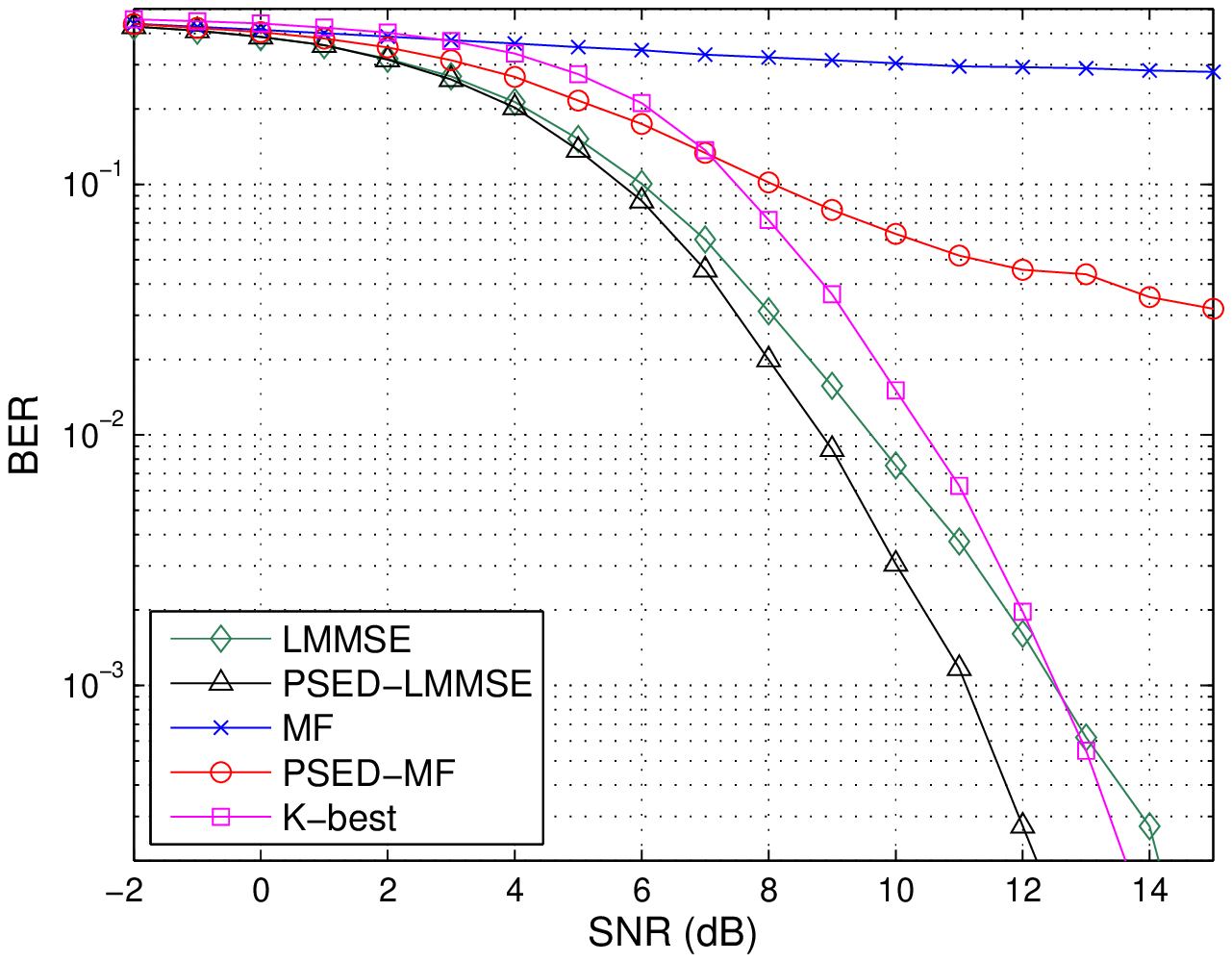}}
   \hspace{-1cm}
  \subfigure[]
   {\includegraphics[width=85mm,height=75mm]{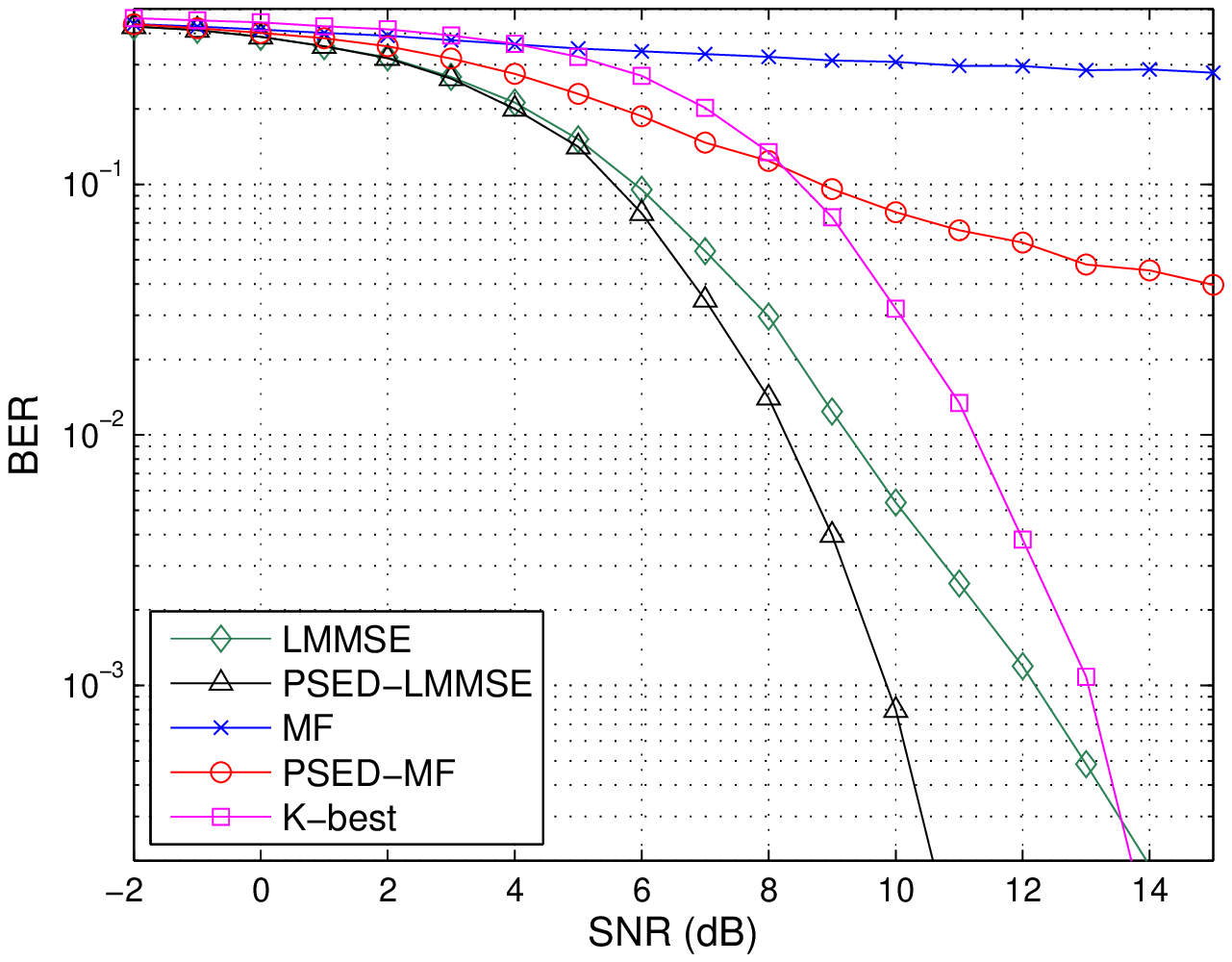}}
     \subfigure[]
  {\includegraphics[width=85mm,height=75mm]{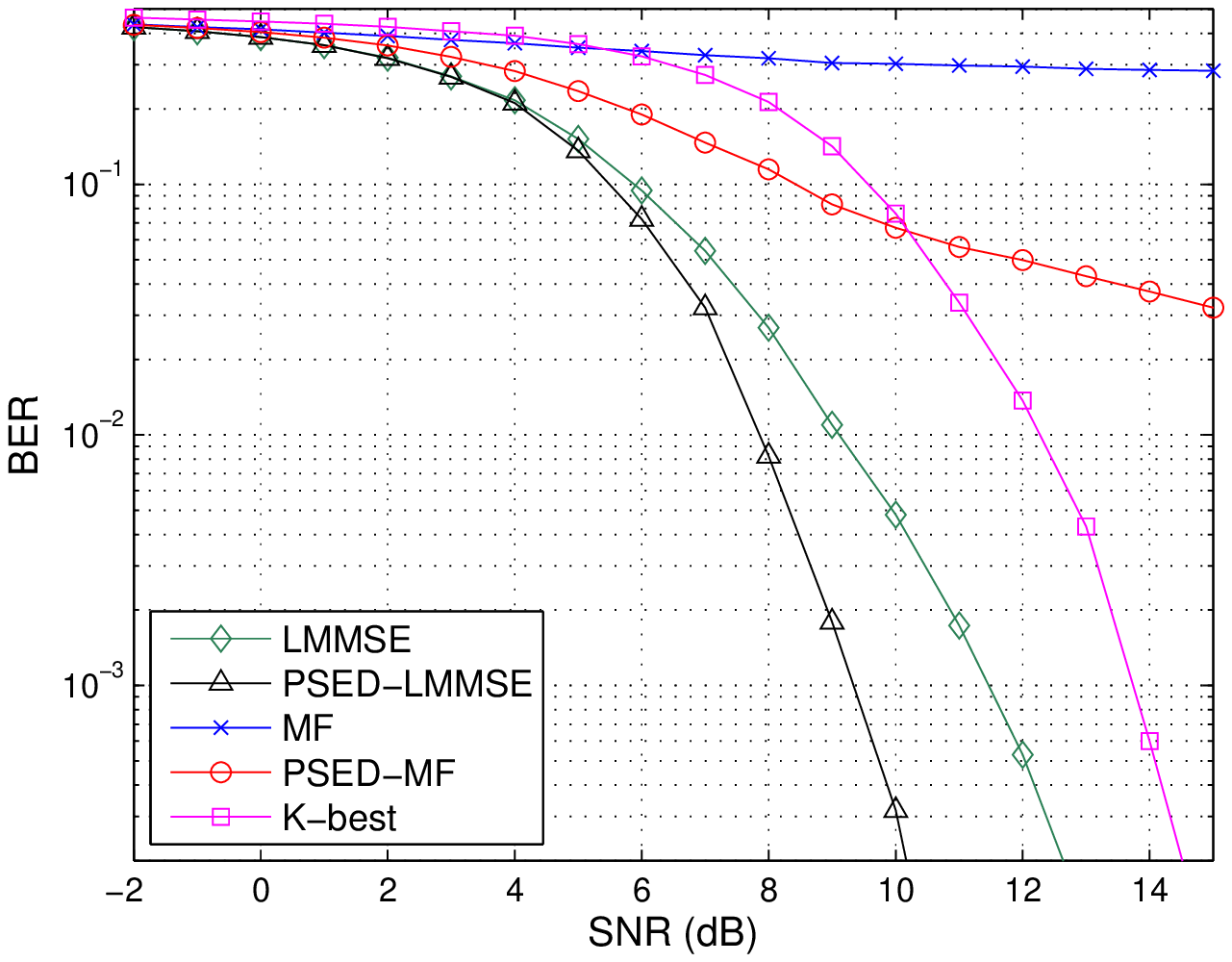}}
       \hspace{-1cm}
  \subfigure[]
   {\includegraphics[width=85mm,height=75mm]{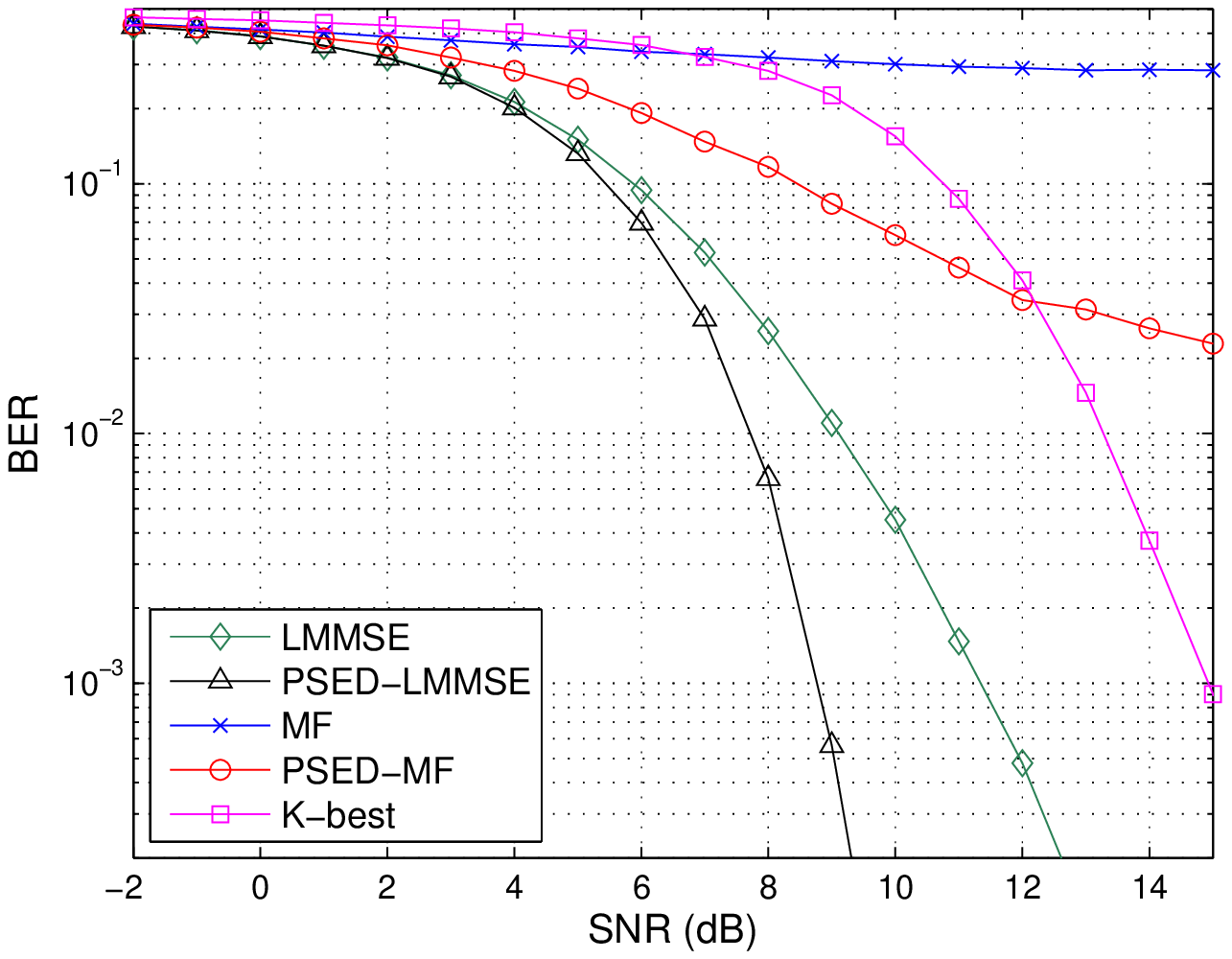}}
  \caption {The BER as a function of SNR for coded systems. (a) $(n_r,n_t)=(32,32)$, (b) $(n_r,n_t)=(64,64)$, (c) $(n_r,n_t)=(128,128)$, and (d) $(n_r,n_t)=(256,256)$.
   } \label{fig:perf_coded}
\end{figure*}

Fig. \ref{fig:perf} shows the symbol error rate (SER) performance of the PSED algorithm and the conventional MIMO detectors.
We consider $4$ different system dimensions, viz., $(n_r,n_t)=(32,32), (64,64), (128,128)$, and $(256,256)$.
We observe that the proposed PSED-LMMSE and PSED-MF outperform their counterparts (LMMSE and MF detectors) by a large margin.
For example, the gain of PSED-LMMSE over the LMMSE detector is more than $6$ dB at $10^{-2}$ SER for all dimensions under test.
We see that with $(n_r,n_t)=(32,32)$, the performance gap of PSED-LMMSE from the optimal ML detector is around $2$ dB.
%
While the performance of PSED-LMMSE is comparable to the $K$-best detector for $(n_r,n_t)=(32,32)$, the gain increases with the dimension, exhibiting $6.5$ dB gain when $(n_r,n_t)=(128,128)$.
We also note that in contrast to PSED-LMMSE, the performance of PSED-MF is not so appealing. This is because the MF detector performs bad  for the SNR range of interest  (i.e., error rate is higher than $15\%$) so that the input error vector after the sparse transformation is non-sparse and the sparse recovery algorithm does not perform well.
In Table \ref{tb:comp2}, we provide the complexity (i.e., the number of complex multiplication) of the detection schemes under consideration. When compared to the complexity of the conventional detector, we see that the complexity required to perform post processing in the PSED detectors is quite small. Note also that in spite of significant performance gain, the complexity of PSED-LMMSE is comparable to that of K-best detector.

We next investigate the performance of detectors for non-square dimensional systems where the dimension of measurement vector is larger than that of transmit vector (e.g., massive MIMO uplink scenario). In this simulations,
we set $n_t = 32$ and check the performance when $n_r$ is $48$, $64$, and $128$ (we already plotted results for $(n_r,n_t)=(32,32)$ in Fig. \ref{fig:perf}).
%
%
We observe that the performance of PSED-MF improves rapidly as $n_r$ increases, closing the gap between PSED-LMMSE and PSED-MF. Indeed, when $(n_r,n_t)=(128,32)$, the performance of PSED-MF is almost identical to that of PSED-LMMSE.
This results is desirable, since the computational complexity of PSED-MF is significantly smaller than that of PSED-LMMSE and LMMSE (see Section \ref{sec:complexity}) so that we make the most of the PSED algorithm with only small computational cost.

In order to see that the gain of PSED algorithm is maintained even for the coded system, we evaluate the performance of the detectors when the forward error correction code (FEC) is employed. We use the convolutional channel code with generator polynomial $(171,131)$. The size of code block is set to 2048 bits and a random interleaver is used between the channel encoder and the symbol modulator. In performing the channel decoding, Viterbi decoder is employed \cite{viterbi}.
In Fig. \ref{fig:perf_coded}, we plot the coded bit error rate (BER) measured as a function of SNR.
We observe that the performance gain of PSED-LMMSE over the conventional detection schemes is well maintained and also increases with the number of antennas.

\begin{figure*}
 \centering
  \subfigure[]
  {\includegraphics[width=110mm,height=85mm]{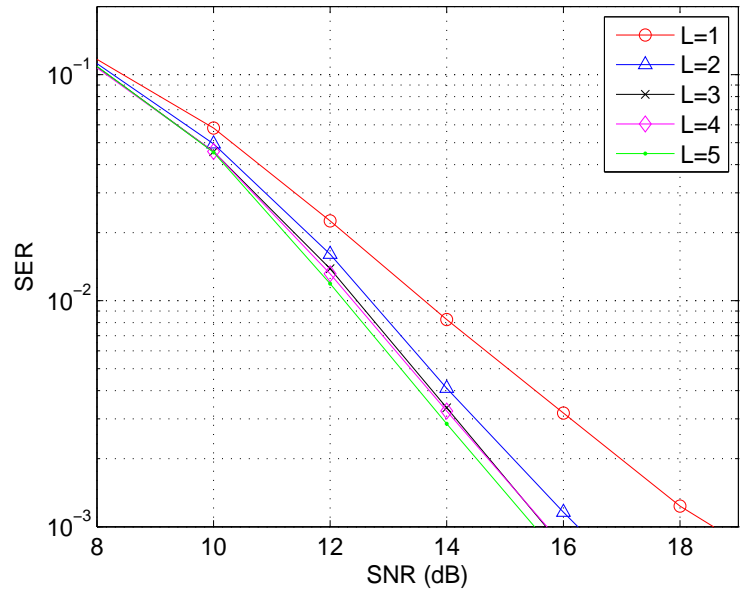}}
  \subfigure[]
   {\includegraphics[width=110mm,height=85mm]{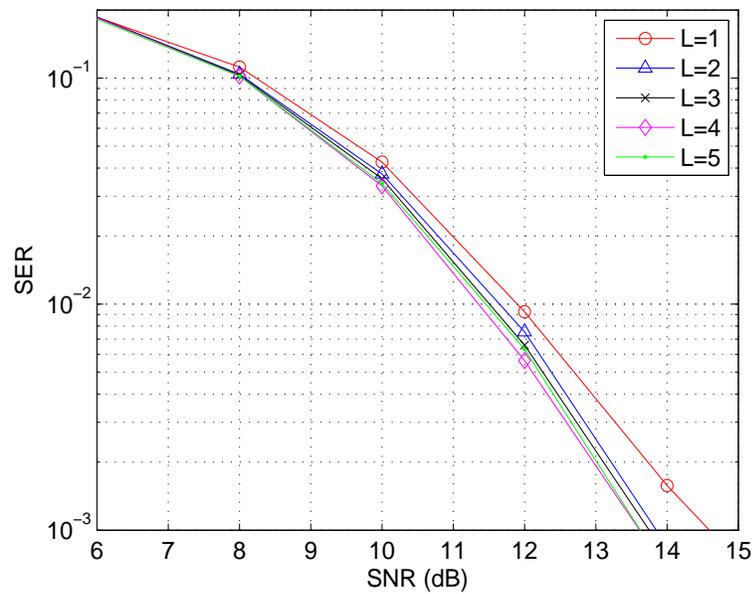}}

  \caption {The SER vs. SNR plot of the PSED-MMSE detector for various choices of $L$: a) $n_r=n_t=16$ and b) $n_r=n_t=32$.
   } \label{fig:ncand}
\end{figure*}

Since the tree branching parameter $L$ of MMP affects computational complexity of the PSED, it is of interest to investigate the impact of $L$ on the performance. Towards this end, we examine the performance of PSED for different choices of $L$ ($L = 1, \cdots, 5$).
As shown in Fig. \ref{fig:ncand}, we observe the meaningful gain from $L=1$ to $L=2$ (note that if $L=1$, then MMP returns to the OMP algorithm) but the gain achieved by further increasing $L$ is pretty marginal.

%
%

\section{Conclusion}	
\label{sec:conclusion}
In recent years, compressive sensing has received much attention in wireless communication industry. However, not much work is available for the information vector detection mainly because the symbol vectors being transmitted are non-sparse. In this paper, we introduced new detection approach exploiting the compressive sensing principle to improve the detection quality of symbol vectors in large-scale wireless communication systems.
Our approach operates in two steps. In the first step, we transform the conventional communication system into the system whose input is the sparse error vector. This mission is accomplished by the conventional linear detection followed by the symbol quantization. For the transformed measurement vector, we next apply the sparse error recovery algorithm followed by the error cancellation to obtain the refined estimate of the transmit symbol vector. In a nutshell, our approach is simple to implement with relatively small computational cost, yet offers substantial gain in performance. Indeed, we observed from the asymptotic performance analysis and empirical simulations that the proposed PSED algorithm achieves significant gain in terms of mean square error (MSE) and symbol error rate (SER) over the conventional detection schemes.


\appendices

\section{Proof that the output streams of the linear MMSE detector are asymptotically uncorrelated }
\label{apx:uncor}
The $i$th output stream of the linear MMSE detector is given by $\hat{s}_i=\mathbf{h}_{i}^{H}\left(\mathbf{H}\mathbf{H}^{H}+\frac{1}{\rm SNR}\mathbf{I}\right)^{-1}$.
The correlation between the $i$th and $j$th output streams is given by
\begin{equation}
E[\hat{s}_i \hat{s}_j^*] = P \mathbf{h}_{i}^{H}\left(\mathbf{H}\mathbf{H}^{H}+\frac{1}{\rm SNR}\mathbf{I}\right)^{-1}\mathbf{h}_{j},
\end{equation}
where $i\neq j$. If we use a matrix inversion lemma $\mathbf{x}^{H}(\mathbf{A}+ \tau \mathbf{x}\mathbf{x}^{H})^{-1} =\frac{\mathbf{x}^{H}\mathbf{A}^{-1}}{1+\tau \mathbf{x}^{H} \mathbf{A}^{-1}\mathbf{x}}$, we can show that
\begin{align}
E[\hat{s}_i \hat{s}_j^*] = \frac{\mathbf{h}_{i}^{H}\left(\mathbf{H}_{[i,j]}\mathbf{H}_{[i,j]}^{H} + \frac{1}{\rm SNR}\mathbf{I}\right)^{-1}\mathbf{h}_{j}}{\left(1+\mathbf{h}_{i}^{H} \left(\mathbf{H}_{[i]}\mathbf{H}_{[i]}^{H} + \frac{1}{\rm SNR}\mathbf{I}\right)^{-1} \mathbf{h}_{i}\right)\left(1+\mathbf{h}_{j}^{H} \left(\mathbf{H}_{[j]}\mathbf{H}_{[j]}^{H} + \frac{1}{\rm SNR}\mathbf{I}\right)^{-1} \mathbf{h}_{j}\right)}, \label{eq:corrs}
\end{align}
where $\mathbf{H}_{[\mathcal{A}]}$ is the submatrix of $\mathbf{H}$ with the columns specified by the index set $\mathcal{A}$ are removed.
According to Lemma 4 in \cite{hoydis}, when $n_t, n_r \rightarrow \infty$, the numerator in (\ref{eq:corrs}) converges to zero and the denominator converges to one.

\section{Proof of \eqref{eq:dl1} and \eqref{eq:dl2}}
\label{apx:conv}
As mentioned, the distribution of $| E |$ is approximated by $\mathcal{N}(n_t P_{e},n_t P_{e} (1-P_{e}))$.
Thus, the quantity $\frac{ | E | }{ n_t }$ follows $\mathcal{N}(P_{e}, \frac{ P_{e} (1-P_{e}) }{ n_t })$ and
\begin{align}
Pr\left(\left|\frac{ |E| }{ n_t } - P_{e}\right|>\epsilon\right) &= 2 Q\left(\frac{\epsilon}{\sqrt{ \frac{ P_{e} (1-P_{e}) }{ n_t }  }} \right) \\
& <   2 \exp\left(- \frac{  \epsilon^{2} n_t^{2}}{2P_{e} (1-P_{e})}\right), \label{eq:vvv}
\end{align}
for any positive $\epsilon$.
When $n_t \rightarrow \infty$, \eqref{eq:vvv} goes to zero and thus $\frac{ | E | }{ n_r } (= \frac{ \beta | E | }{n_t })$ also converges to $P_{e}\beta$.

\end{document}